\documentclass[aps,prb,amsmath,amssymb,floatfix,twocolumn,10pt]{revtex4-2}
\usepackage{graphicx}% Include figure files
\usepackage[colorlinks,allcolors=blue]{hyperref}
\usepackage{braket}
\usepackage{bm}% bold math
\usepackage{comment}
\usepackage{xcolor}
\usepackage{soul}
\setstcolor{red}
\usepackage{caption}
\usepackage{booktabs}

\usepackage{ragged2e}
\DeclareCaptionJustification{justified}{\justifying}
\usepackage[caption=false]{subfig}
\DeclareCaptionOption{justified}[]{\caption@setjustification{justified}}

\DeclareMathAlphabet\mathbfcal{OMS}{cmsy}{b}{n}

\def\YBCO248{YBa$_2$Cu$_4$O$_8$}

\def\BSCCOs{Bi$_2$Sr$_2$CaCu$_2$O$_{8+x}$ }

\def\bnabla{\bm{\nabla}}

\def\br{{\bf r}}

\def\bu{{\bm u}}
\def\bw{{\bm w}}

\def\bk{{\bf k}}
\def\bq{{\bf q}}

\def\bJ{{\bf J}}

\def\bR{{\bf R}}

\def\vr{\vec{r}}
\def\qf{q_\phi}

\newcommand\redsout{\bgroup\markoverwith{\textcolor{red}{\rule[0.5ex]{2pt}{0.4pt}}}\ULon}

\makeatletter
\newsavebox{\@brx}
\newcommand{\llangle}[1][]{\savebox{\@brx}{\(\m@th{#1\langle}\)}%
  \mathopen{\copy\@brx\kern-0.5\wd\@brx\usebox{\@brx}}}
\newcommand{\rrangle}[1][]{\savebox{\@brx}{\(\m@th{#1\rangle}\)}%
  \mathclose{\copy\@brx\kern-0.5\wd\@brx\usebox{\@brx}}}
\makeatother

\begin{document}

\title{Signatures of Gaussian superconducting fluctuations in nonlocal noise magnetometry}

\author{Dror~Orgad}
\affiliation{Racah Institute of Physics, The Hebrew University,
  Jerusalem 91904, Israel}

\date{\today}

\begin{abstract}
%We calculate the contribution of Gaussian superconducting fluctuations to the two-point magnetic noise 
%spectrum, which can be measured by a pair of spin qubits such as nitrogen vacancy centers in a diamond. 
%We do so using the time-dependent Ginzburg-Landau theory, concentrating on the case of a two-dimensional 
%system in equilibrium and under the influence of a uniform electric field. 

We calculate the two-point magnetic noise spectrum arising from Gaussian superconducting fluctuations, 
a quantity directly measurable by spin qubit pairs such as nitrogen vacancy centers in diamond. The analysis 
utilizes the time-dependent Ginzburg-Landau theory, reflecting the direct contribution of fluctuating 
Cooper pairs to the current correlations and consequent magnetic noise. We focus on two-dimensional 
systems and consider them both in equilibrium and under a uniform electric field.
The signal is expected to be particularly pronounced in high-temperature superconductors, and we contrast 
our findings with the predicted signatures of a vortex liquid to offer an additional route to 
elucidate the nature of fluctuations in these systems. For completeness and comparison, we also present  
analogous results for one-dimensional wires. 
%our findings offer a way to 
%further elucidate the nature of fluctuations in these systems.
%The signal is expected to be strongest in high-temperature superconductors, and our findings offer a 
\end{abstract}

\maketitle

\section{Introduction}

The effects of superconducting fluctuations above the transition temperature $T_c$ have been the focus of 
intense research over the years \cite{LarkinVarlamov}. Such fluctuations may be enhanced by disorder or 
inhomogeneity  \cite{Aslamasov1980,Skvortsov2005,Petrovic2016,Zhao-disorder-NbSe2,Liu2024}, by proximity 
to a quantum critical point \cite{Saxena2000,Shibauchi-Pnictides,Sachdev2010,Lederer-nematic,Wang2016}, 
and in systems with short coherence length or small superfluid density \cite{Mohit1992,Emery-Kivelson}. 
The cuprate high-temperature superconductors exhibit many of these enhancing factors, especially when underdoped \cite{OurReview}.  %especially in the 
%underdoped region of their phase diagram \cite{OurReview}. 
Therefore, it is not surprising that large 
superconducting fluctuations have been observed in these materials by multiple probes, including paraconductivity 
\cite{Popcevic2018}, diamagnetism \cite{Ong-diamagnetism,YBCO-Gaussian} and the Nernst effect 
\cite{Ong-Nernst,Chang-Gaussian}. However, the precise nature of the fluctuations is not yet settled.
One possibility is that they derive from a vortex liquid state 
%that persists at temperatures significantly above $T_c$ 
\cite{Ong-diamagnetism,Ong-Nernst,Podolsky-PRL,Gideon-PRB}. 
In this scenario, superconductivity is suppressed by phase fluctuations of the order parameter, while 
its amplitude remains large well above $T_c$. Alternatively, the measured 
signals may emerge from fluctuations of the order parameter that also involve its amplitude, 
possibly in the form of Gaussian fluctuations \cite{YBCO-Gaussian,Chang-Gaussian,Ido-Nernst,Caprara-paraconductivity}.

Measurement of both equilibrium and non-equilibrium noise has become a valuable technique for probing 
electronic systems, revealing key properties such as temperature, linear response, and the charge of 
elementary excitations \cite{Johnson,Nyquist,Buttiker,Reznikov1999,Blanter2000,Noise-localcorr}. 
This also holds true for systems at the edge or within a superconducting phase 
\cite{Nagaev,Weissman,Dolgirev22,Chatterjee22,Levchenko-SCnoise,Curtis24,Liu2025,Chatterjee26}. Much of the renewed interest 
in noise measurements stems from advances over the past decade in using nitrogen vacancy (NV) 
centers in diamonds 
to detect weak fluctuating magnetic fields \cite{Rovny2024,Maletinsky2012,Kolkowitz2015,Boss2017, Glenn2018,
Andersen2019,Rovny-covariance,deLeon-multiplexed,Kolkowitz-multiplexed,Le-subdiffraction,Degen-subtip}. 
Recently, simultaneous optical readout of multiple NV centers was demonstrated both above 
\cite{Rovny-covariance,deLeon-multiplexed,Kolkowitz-multiplexed} and below \cite{Le-subdiffraction,Degen-subtip,deLeon-subdiff} 
the diffraction limit. Consequently, spatial correlations of noise, now experimentally accessible, have been 
proposed as a tool for characterizing the non-equilibrium distribution function in current-biased conductors 
\cite{Sarang-spatial}. In the present study, we consider the equilibrium and non-equilibrium contributions of 
Gaussian superconducting fluctuations to spatially resolved magnetic noise correlations, aiming to identify 
testable signatures of such fluctuations.

To this end, we employ the stochastic time-dependent Ginzburg-Landau (TDGL) formalism, reviewed in 
Sec. \ref{sec:formalism}, to compute the direct contribution of fluctuating Cooper pairs to the 
transverse-current correlations in a two-dimensional system. Beyond these Aslamazov-Larkin (AL) 
processes \cite{Aslamazov1968}, additional contributions exist that are not captured by the TDGL 
(see however Ref. \cite{TDGL-Alex}). They arise from the scattering of electrons off fluctuating 
pairs and from induced fluctuations in the normal electrons' density of states. Notably, 
the scattering contribution, known as the Maki-Thompson (MT) correction 
\cite{Maki1,Maki2,Thompson1,Thompson2}, is dominant for a two-dimensional $s$-wave 
superconductor under weak pair-breaking conditions \cite{LarkinVarlamov}. Nevertheless, 
the experimentally observed paraconductivity in the cuprates is largely explained by AL processes 
\cite{LarkinVarlamov}, suggesting that $d$-wave pairing or significant pair-breaking effects 
suppress the MT contribution in these materials.

Our results, detailed in Sec. \ref{sec:results}, extend the findings of Ref. \cite{Curtis24} regarding 
the AL contribution to magnetic noise by considering both nonlocal measurements and driven non-equilibrium 
systems. As such, they provide information about the wavevector structure of the underlying 
current correlations and the way they are affected by a transport current.  
Specifically, we find that in equilibrium, the two-point magnetic-noise spectrum from 
Gaussian superconducting fluctuations is essentially frequency-independent across the NV-center sensing 
band. Its dependence on the combined height of the sensors above the sample $z_c$, and on their lateral 
separation $\Delta r$, exhibits non-monotonic dependence on $z_c$ at large $\Delta r$, and 
characteristic power-law asymptotic decays in $z_c$ and $\Delta r$. These are accompanied by 
power-law divergences as a function of $T-T_c$. The linear response of the noise to a driving 
electric field $E$ vanishes under particle-hole symmetry. With (typically weak) particle-hole 
asymmetry the response becomes odd in frequency and spatially anisotropic, following 
$\bm E\cdot\bm{\Delta r}=E\Delta r\cos\theta$. Being proportional to $\Delta r$ it cannot 
be detected by a local measurement. The $O(E^2)$ contribution, on the other hand, is present even 
in the particle-hole symmetric case, persists down to zero frequency, and features 
both isotropic and $\cos 2\theta$ quadrupolar components. 

To make further use of the above results, we contrast them with the predicted signal from a vortex liquid 
above its Berezinskii-Kosterlitz-Thouless (BKT) transition. For this purpose, we develop in 
Sec. \ref{sec:vortexfluct} a stochastic coarse-grained framework for the dynamics of free vortices 
and use it to calculate their vorticity correlations and magnetic noise. In equilibrium, the Gaussian 
and vortex signals share a similar far-field spatial dependence, but are controlled by different length 
and time scales: the Gaussian crossover is set by the coherence length and Cooper-pair 
relaxation rate, whereas the vortex signal is governed by the BKT correlation length 
and the much smaller screening relaxation rate. Their driven responses provide further distinctions. 
In the particle-hole symmetric case, opposite drift of vortices and antivortices couples fluctuations 
of the total free-vortex density to the vorticity and produces an $O(E^2)$ response. At long wavelengths 
this conversion enhances the zero-frequency noise, in contrast to the suppression found for Gaussian fluctuations, while on scales shorter than the vortex screening length the drift-induced relaxation of vorticity fluctuations generates a $-\cos 2\theta$ quadrupolar component, corresponding to enhanced correlations for probes separated perpendicular to the current. At stronger currents, 
current-induced vortex-antivortex unbinding provides an additional source of free vortices 
and can further enhance this short-distance anisotropic response, although the concomitant 
reduction of the screening length eventually modifies it. Thus, the characteristic frequency, 
spatial crossover scale, sign of the isotropic nonlinear response, and orientation of the 
quadrupolar signal offer complementary means of distinguishing vortex fluctuations 
from Gaussian order-parameter fluctuations. In Sec. \ref{sec:exp} we discuss the expected 
magnitude and experimental accessibility of both Gaussian and vortex signatures in representative 
quasi-two-dimensional superconductors.

Finally, for completeness and comparison, we include an appendix with a similar analysis for a 
one-dimensional superconducting wire. 
%For completeness, we also present a similar set of calculations for 
%%the two-point magnetic noise spectrum due to superconducting fluctuations in 
%a one-dimensional superconducting wire. 
The theory does not take into account 
quantum and thermal phase slips that are expected to play an important role in the physical systems 
\cite{Zaikin-review}. However, it may be relevant to a limited range of temperatures above but not too 
close to $T_c$, as revealed by experiments \cite{Dynes,Mitra-1d}.

\section{Formalism}
\label{sec:formalism}

The TDGL equation may be viewed as a model for the critical dynamics of a charge $-2e$ superconducting 
order parameter $\psi(\br,t)$ \cite{LarkinVarlamov,Hohenberg77}
\begin{equation}
\label{eq:TDGL}
    (\tau+i\tau')\left[\frac{\partial}{\partial t}-2ie\Phi(\br,t)\right]\psi(\br,t)=
    -\frac{\delta{\cal F}}{\delta\psi^*(\br,t)}+\zeta(\br,t),
\end{equation}
where henceforth $\hbar=c=k_B=1$. The deterministic dynamics is governed by the Ginzburg-Landau free energy  
\begin{equation}
\label{eq:GLF}
{\cal F}=\int d\br\left[ a|\psi|^2+\frac{b}{2}|\psi|^4+\frac{1}{2m^*}|\bnabla \psi|^2\right],
\end{equation}
while thermodynamic fluctuations are introduced through a Gaussian delta-correlated Langevin force  $\zeta(\br,t)$. We work throughout in the gauge $\bm A=0$. Gauge invariance then dictates the coupling 
of the scalar electric potential $\Phi(\br,t)$ in the time derivative term. The dimensionless kinetic 
coefficients $\tau$ and $\tau'$ describe the dissipative and reactive response of the system, 
respectively. Particle-hole symmetry enforces $\tau'=0$ \cite{Dorsey91}, and typically 
$\eta\equiv\tau'/\tau\sim T_c/E_F\ll 1$ \cite{Ebisawa71}. 
Finally, the fluctuation-dissipation theorem constrains the Langevin force correlations to be 
\begin{equation}
\label{eq:zetacorr}
\langle\zeta^*(\br,t)\zeta(\br',t')\rangle=2\tau T\delta(\br-\br')\delta(t-t').
\end{equation}
The coefficients of the model, including $a=a_0(T-T_c)$, are ultimately to be determined by experiments. 
However, they can also be derived from the microscopic theory, which yields $\tau=\pi a_0/8$ \cite{LarkinVarlamov}. 

\begin{figure}[t!!!]
\centering
\includegraphics[width=\columnwidth,clip=true]{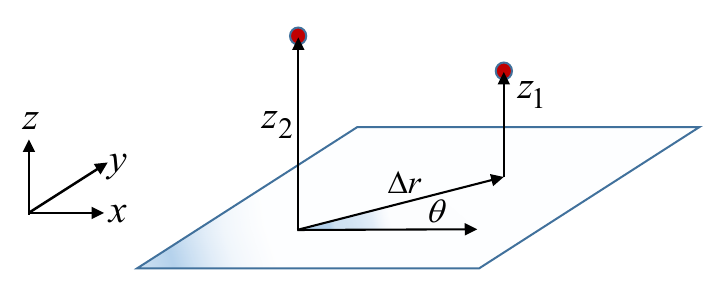}
\vspace{0pt}
\caption{The geometry of the two-dimensional system. The red dots indicate the positions of the NV centers 
used for the magnetic sensing. When an external current is present, it flows in the $x$ direction.}
\label{fig:geometry2d}
\end{figure}

For $T>T_c$ and outside the critical Ginzburg region, the Gaussian 
approximation amounts to discarding the quartic term in 
Eq. (\ref{eq:GLF}). This leads to a linear TDGL equation whose Fourier-space solution in $d$ dimensions is given in 
terms of its Green's function according to 
\begin{equation}
\psi(\bk,\omega)=\int \frac{d^d k'}{(2\pi)^d} \frac{d\omega'}{2\pi} G(\bk,\omega;\bk' ,\omega')\zeta(\bk',\omega').
\end{equation}
In the absence of an electric potential 
$G(\bk,\omega;\bk' ,\omega')=G_0(\bk,\omega)(2\pi)^{d+1}\delta(\bk-\bk')\delta(\omega-\omega')$ with
\begin{equation}
\label{eq:G0}
G_0(\bk,\omega)=\frac{1}{-(\tau+i\tau')i\omega+k^2/(2m^*)+a}.
\end{equation}

\section{Noise from a two-dimensional system}
\label{sec:results}

Consider a two-dimensional superconductor occupying the $x-y$ plane and sustaining an instantaneous 
electric current distribution $\bJ(\br,t)$, see Fig. \ref{fig:geometry2d}. In the following, bold letters 
denote two-dimensional vectors, while an arrow is used to signify three-dimensional vectors.  
Neglecting retardation effects, the $z$-component of the magnetic field at a point $\vr$ is given 
(in Gaussian units) by \cite{Curtis24}  
\begin{equation}
\label{eq:Bz2d}
B_z(\vr,t)=2\pi i\int\frac{d^2 q}{(2\pi)^2}\frac{d\nu}{2\pi}e^{i(\bq\cdot\br-\nu t)-q|z|}J_\perp(\bq,\nu),
%\times\frac{1}{q}\left[q_xJ_y(\bq,\nu)-q_yJ_x(\bq,\nu)\right].
\end{equation}
where $J_\perp(\bq,\nu)=[\hat\bq\times\bJ(\bq,\nu)]\cdot \hat z$, with $\hat{\bq}=\bq/q$. The transverse current density 
$J_\perp$ originates from regions where lines of constant amplitude are not parallel to lines of constant phase, 
thus leading to a non-zero curl. For example, Gaussian fluctuations may result in such regions with localized 
current loops. These are to be distinguished from vortex configurations whose amplitude is essentially constant 
outside the vortex core. 

For the planar components one obtains 
\begin{equation}
\label{eq:Bplan2d}
\mathbf{B}(\vr,t)=2\pi\,{\rm sgn}(z)\int\frac{d^2 q}{(2\pi)^2}\frac{d\nu}{2\pi}{\hat \bq}e^{i(\bq\cdot\br-\nu t)-q|z|}
J_\perp(\bq,\nu).
\end{equation}

Our main focus is the two-point magnetic noise spectrum 
\begin{equation}
\label{eq:Szz}
S_{zz}(\vr_1,\vr_2,\omega)=\int dt e^{i\omega t}\langle B_z(\vr_1,t) B_z(\vr_2,0)\rangle.  
\end{equation}
This quantity is directly related to phase-accumulation covariance measurements in the case where 
the NV quantization axis is along the $z$ direction \cite{Rovny-covariance,Rovny2024}. 
During a sensing sequence, 
the $i$-th NV center acquires the phase $\phi_i=\gamma_{\rm NV}\int dt\,w_i(t)B_z(\br_i,t)$, where 
$\gamma_{\rm NV}=g\mu_B/\hbar$ and $w_i(t)$ is the sequence-dependent filter function. 
The measured phase covariance is therefore 
\begin{equation}
\label{eq:phasecov}
\langle\phi_1\phi_2\rangle=\gamma_{\rm NV}^2\int\frac{d\omega}{2\pi}W_1(-\omega)W_2(\omega)S_{zz}(\br_1,\br_2,\omega),
\end{equation}
with $W_i(\omega)$ the corresponding filter function in frequency space.

The closely related function
\begin{equation}
\label{eq:Sperp}
S_{+-}(\vr_1,\vr_2,\omega)=\int dt e^{i\omega t}\langle B_+(\vr_1,t) B_-(\vr_2,0)\rangle,  
\end{equation}
which involves the planar components $B_\pm=B_x\pm iB_y$, obeys 
for the translationally invariant cases considered here, 
$S_{+-}(\vr_1,\vr_2,\omega)={\rm sgn}(z_1 z_2)S_{zz}(\vr_1,\vr_2,\omega)$. Its value at the NV 
transition frequency, $\Omega_{\rm NV}$, can be measured by a covariance relaxometry scheme 
\cite{Rovny2024,Le-subdiffraction}, which is sensitive to the correlated 
part of the decay rates $\Gamma_{12}\propto \gamma_{\rm NV}^2 S_{+-}(\br_1,\br_2,\Omega_{\rm NV})$.

Given the dependence of the noise spectrum on current 
correlations, we proceed to examine the contribution that they acquire from Gaussian superconducting fluctuations. 
%which clearly depends on the current correlations in the plane. Consequently, we turn now to calculate the 
%contribution of Gaussian superconducting fluctuations to these correlations and to the resulting magnetic noise. 
We begin by considering a system in equilibrium.

\begin{figure}[t!!!]
\centering
\includegraphics[width=\columnwidth,clip=true]{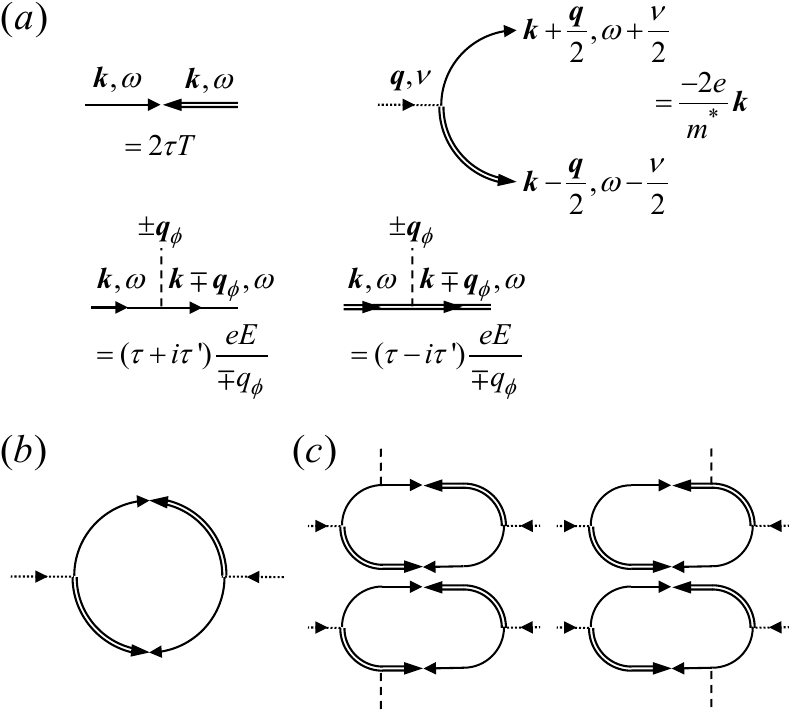}
\captionsetup[subfigure]{labelformat=empty}
    \subfloat[\label{subfig:vertices}]{}
    \subfloat[\label{subfig:JJ}]{}
    \subfloat[\label{subfig:1st}]{}
\caption{(a) The vertices of the theory. Solid lines indicate $G_0(\bk,\omega)$, while double lines stand 
for $G_0^*(\bk,\omega)$. The dashed lines are associated with the electric potential. (b) The diagram for 
the equilibrium current-current correlator. (c) First order diagrams for the current-current correlator.}
\label{fig:vertices}
\end{figure}

\subsection{Equilibrium noise}

The expression for the electric current density due to superconducting fluctuations  
\begin{eqnarray}
\label{eq:Jsup}
\nonumber
\bJ(\bq,\nu)=\frac{-2e}{m^*}\int \frac{d^2k}{(2\pi)^2}\frac{d\omega}{2\pi}&&\bk\,\psi^*\left(\bk-\frac{\bq}{2},\omega-\frac{\nu}{2}\right)\\
\times&&\psi\left(\bk+\frac{\bq}{2},\omega+\frac{\nu}{2}\right),
\end{eqnarray}
and Eqs. (\ref{eq:zetacorr})-(\ref{eq:G0}) lead to the current-current correlator, whose diagrammatic representation 
is depicted in Fig. \ref{subfig:JJ}. Analytically, the transverse-current correlation function, relevant for our purposes, 
takes the form
\begin{eqnarray}
\nonumber
\langle J_\perp(\bq,\nu)J_\perp(\bq',\nu')\rangle_{eq}  &=& -\frac{\pi^3 e^2T^2}{T-T_c}I_{eq}(\bar q,\bar\nu)\\
&&\times\delta(\bq+\bq')\delta(\nu+\nu'),
\end{eqnarray}
where for $\eta=0$ the kernel becomes 
%For $\eta=0$ the kernel becomes 
\begin{eqnarray}
\label{eq:I2deq}
\nonumber
\!\!\!\!\!\!\!\!\!\!I_{eq}(\bar q,\bar \nu)&=&-\frac{1}{2\bar q^2}\ln\left[(1+\bar q^2)^2+4\bar\nu^2\right]\\
\nonumber
&-&\frac{1}{2\bar q^2}{\rm Re}\left\{\frac{\sqrt{\bar q^2+(\bar q^2-i\bar \nu)^2}}{i\bar\nu} \right.\\
\nonumber&\times&\left[{2\,{\rm arctanh}\left(\frac{\bar q^4+\bar q^2(1-i\bar\nu)+i\bar\nu}
{(1+\bar q^2)\sqrt{\bar q^2+(\bar q^2-i\bar\nu)^2}} \right)} \right. \\
&+&\left.\left. \ln\left(\frac{\bar q^2-i\bar\nu+\sqrt{\bar q^2+(\bar q^2-i\bar\nu)^2}}
{-\bar q^2+i\bar\nu+\sqrt{\bar q^2+(\bar q^2-i\bar\nu)^2}}   \right)\right]\right\}.
\end{eqnarray}
In the derivation, we have adopted the microscopic relation $\tau=\pi a_0/8$, and introduced 
the dimensionless quantities $\bar q=\xi q/2$ and $\bar\nu=\nu/\omega_0$. Here, $\xi(T)=1/\sqrt{2m^*a}$ denotes 
the (temperature dependent) coherence length, and $\omega_0(T)=4a/\tau$ is the characteristic frequency scale 
associated with the inverse lifetime of the Cooper pairs.
%For $\eta=0$ the kernel becomes 
%\begin{eqnarray}
%\label{eq:I2deq}
%\nonumber
%\!\!\!\!\!\!\!\!\!\!I_{eq}(\bar q,\bar \nu)&=&-\frac{1}{2\bar q^2}\ln\left[(1+\bar q^2)^2+4\bar\nu^2\right]\\
%\nonumber
%&-&\frac{1}{2\bar q^2}{\rm Re}\left\{\frac{\sqrt{\bar q^2+(\bar q^2-i\bar \nu)^2}}{i\bar\nu} \right.\\
%\nonumber&\times&\left[{2\,{\rm arctanh}\left(\frac{\bar q^4+\bar q^2(1-i\bar\nu)+i\bar\nu}
%{(1+\bar q^2)\sqrt{\bar q^2+(\bar q^2-i\bar\nu)^2}} \right)} \right. \\
%&+&\left.\left. \ln\left(\frac{\bar q^2-i\bar\nu+\sqrt{\bar q^2+(\bar q^2-i\bar\nu)^2}}
%{-\bar q^2+i\bar\nu+\sqrt{\bar q^2+(\bar q^2-i\bar\nu)^2}}   \right)\right]\right\},
%\end{eqnarray}
%when expressed in terms of the dimensionless quantities $\bar q=\xi q/2$ and $\bar\nu=\nu/\omega_0$. 
%Here, $\xi(T)=1/\sqrt{2m^*a}$ denotes the (temperature dependent) coherence length, and $\omega_0(T)=4a/\tau$ 
%is the characteristic frequency scale associated with the inverse lifetime of the Cooper pairs

Defining $\bar\omega=\omega/\omega_0(T)$, we subsequently obtain for the magnetic noise spectrum
\begin{eqnarray}
\label{eq:S2deq}
\nonumber
\!\!\!S^{eq}_{zz}(\vr_1,\vr_2,\omega)= S_0\frac{T^2}{T_c^2}\int_0^\infty d\bar q \,\bar q\, e^{-2\bar q z_c/\xi} \\
\times J_0(2\bar q\Delta r/\xi)
I_{eq}(\bar q,\bar\omega),
\end{eqnarray}
where $\Delta r=|\br_1-\br_2|$, and $J_0$ is the Bessel function. In Eq. (\ref{eq:S2deq}), we introduced the combined distance 
\begin{equation}
z_c=|z_1|+|z_2|,
\end{equation} 
of the two field-sensing impurities from the plane, and 
\begin{equation}
\label{eq:S0def}
S_0=\frac{\pi e^2 T_c}{\xi_0^2},
\end{equation}
which sets the overall noise scale. In addition, we defined 
\begin{equation}
\xi_0=\frac{1}{\sqrt{2m^*a_0T_c}},
\end{equation}
which in a clean two-dimensional system is related to the BCS zero-temperature 
coherence length $\xi_{0,\rm BCS}=\hbar v_F/\pi\Delta_0$ via $\xi_0=0.9\xi_{0,\rm BCS}$.
In the dirty limit $\xi_0=\sqrt{\xi_{0,\rm BCS}\ell}$, where $\ell$ is the mean free path 
\cite{LarkinVarlamov}. 
Finally, to bring out the temperature dependence near $T_c$ we also introduce the dimensionless temperature difference
\begin{equation}
\varepsilon=\frac{T-T_c}{T_c},
\end{equation}
in terms of which $\xi(T)=\xi_0/\sqrt{\varepsilon}$, and 
\begin{equation}
\omega_0(T)=\frac{32}{\pi}\varepsilon T_c.
\end{equation}

\begin{figure}[t!!!]
\centering
\includegraphics[width=\columnwidth,clip=true]{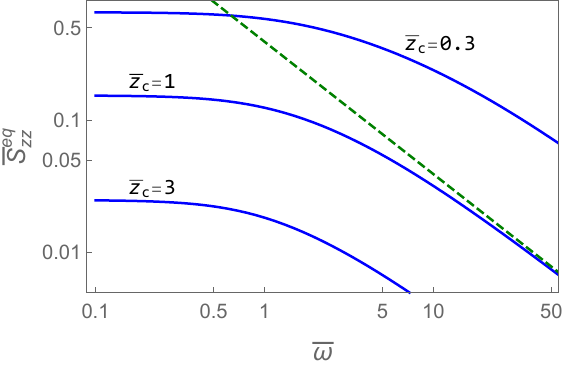}
\vspace{10pt}
\caption{$\bar S_{zz}^{eq}=S_{zz}^{eq}/[(T/T_c)^2S_0]$ as a function of $\bar\omega$ 
for various values of $\bar z_c=\sqrt{\varepsilon}z_c/\xi_0$ and $\Delta r=0$. The dashed line depicts a 
$\bar\omega^{-1}$ behavior.}
\label{fig:S2omega}
\end{figure}

Equations (\ref{eq:I2deq}) and (\ref{eq:S2deq}) allow for a numerical determination of the full 
frequency-dependent noise spectrum for arbitrary sensor positions. Figure \ref{fig:S2omega} illustrates this 
dependence for several values of $z_c$. The noise spectral density remains nearly constant up to a 
characteristic crossover frequency, beyond which it exhibits an asymptotic $\omega^{-1}$ power-law decay, 
as confirmed by a large-frequency expansion of the kernel in Eq. (\ref{eq:I2deq}). Below we demonstrate analytically 
that for large $z_c$ this crossover occurs at $\omega_0(T)$. It shifts to higher frequencies as either $z_c$ or $\Delta r$ 
are decreased. Given that $\omega_0/2\pi\simeq 0.2$~THz for $T-T_c=1$K, well above the experimentally 
relevant NV sensing frequencies \cite{Rovny2024},
%which is about two orders of magnitude larger than the highest frequency accessible to NV sensing \cite{Rovny2024}, 
we focus in the following on the limit $\omega=0$.

Figure \ref{fig:S2drz} shows the dependence of the magnetic noise amplitude on the NV center positions. At large distances, 
the amplitude decays asymptotically following a power law, scaling as $z_c^{-2}$ and $\Delta r^{-3}$ for large $z_c$ and $\Delta r$, 
respectively. Also notable is the non-monotonic dependence on $z_c$ for large $\Delta r$.  We can analytically derive this behavior, 
as well as the aforementioned frequency dependence, for a couple of scenarios, which we explore next.

\subsubsection{$\Delta r=0$, $\omega=0$}
At zero-frequency the equilibrium kernel reduces to 
\begin{equation}
I_{eq}(\bar q,0)=\frac{1}{\bar q^2}\left[\frac{2\bar q\,{\rm arcsinh}(\bar q)}{\sqrt{1+\bar q^2}} -\ln(1+\bar q^2)\right].
\end{equation}
To a good approximation, this exact result may be replaced by the Lorentzian $\ln 4/[\ln 4+\bar q^2]$ and used 
in Eq. (\ref{eq:S2deq}) to yield
%When $|z|\gg\xi$ the integral in Eq. (\ref{eq:S2deq}) is dominated by the regime $\bar q\ll 1$ for which $I_{eq}(\bar q,0)=1$. 
%For $|z|\ll \xi$ the integral receives a contribution also from $\bar q\gg1$ where $I_{eq}(\bar q,0)=\ln 4/\bar q^2$. 
%Applying these approximations we arrive at 
\begin{eqnarray}
\label{eq:Seqdr0omega0}
\nonumber
\hspace{-0.5cm}S^{eq}_{zz}(\vr_1,\vr_2,0)&&\\
\nonumber
&&\hspace{-1cm}=-S_0\frac{T^2}{T_c^2}\ln 4\left\{\cos\left(2\sqrt{\ln 4}z_c\right){\rm Ci}\left(2\sqrt{\ln 4}z_c\right)\right.    \\
\nonumber
&&\hspace{-1cm}+\left. \sin\left(2\sqrt{\ln 4}z_c\right)\left[{\rm Si}\left(2\sqrt{\ln 4}z_c\right)-\pi/2\right]\right\} \\
&&\hspace{-1cm}=S_0\frac{T^2}{T_c^2}\left\{\renewcommand{\arraystretch}{2.4} 
\begin{array}{cc} 
 \displaystyle{-2-\ln 4\ln\left(\frac{\sqrt{\varepsilon}z_c}{\xi_0}\right)} & 
\displaystyle{z_c\ll\frac{\xi_0}{\sqrt{\varepsilon}}} \\
\displaystyle{\frac{1}{4\varepsilon}\frac{\xi_0^2}{z_c^2}} & \displaystyle{z_c\gg\frac{\xi_0}{\sqrt\varepsilon}}
\end{array}\right.
\renewcommand{\arraystretch}{1} 
,
\end{eqnarray}
where Ci and Si are the cosine and sine integral functions. A similar result was obtained 
in Ref. \cite{Curtis24}.

As a general remark, let us note that the limit $z_c\ll\xi$ implies $\varepsilon\ll (\xi_0/z_c)^2$. Hence it may be relevant only 
if the temperature is very close to $T_c$, where the underlying Gaussian approximation is no longer valid. 
We discuss this point in the context of the experimental systems in Sec. \ref{sec:exp}.

\subsubsection{$z_c\gg \xi_0/\sqrt{\varepsilon}$}
\begin{figure}[t!!!]
\centering
\includegraphics[width=\columnwidth,clip=true]{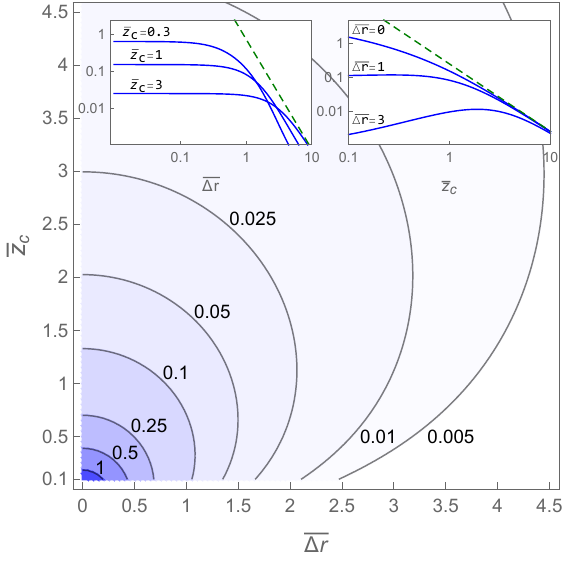}
\vspace{10pt}
\caption{Contour plot of $S_{zz}^{eq}(\omega=0)/[(T/T_c)^2S_0]$ 
as a function of $\bar z_c=\sqrt{\varepsilon}z_c/\xi_0$ and $\overline{\Delta r}=\sqrt{\varepsilon}\Delta r /\xi_0$. 
The left inset includes cuts as a function of $\overline{\Delta r}$, where the dashed line depicts a $\overline{\Delta r}^{\,-3}$ decay. 
The right inset shows cuts as a function of $\bar z_c$, where the dashed line represents a $\bar z_c^{\,-2}$ decay.}
\label{fig:S2drz}
\end{figure}

When at least one of the NV centers is located a distance greater than $\xi$ from the plane, the integral in Eq. (\ref{eq:S2deq}) 
is dominated by the regime $\bar q\ll 1$ and we may replace $I_{eq}(\bar q,\bar\nu)$ by
\begin{equation}
I_{eq}(\bar q\rightarrow 0,\bar\nu)=\frac{\pi}{2|\bar\nu|}-\frac{\rm arccot(2\bar\nu)}{\bar\nu}-\frac{\ln(1+4\bar\nu^2)}{4\bar\nu^2}.
\end{equation}
Consequently, we find
\begin{eqnarray}
\label{eq:Seqlz}
\nonumber
\!\!\!\!\!\!\!\!\!\!\!\!S^{eq}_{zz}(\vr_1,\vr_2,\omega)&=& \frac{S_0}{4}\frac{T^2}{T_c^2}\frac{\xi_0^2}
{\varepsilon}\frac{z_c}{\left[\Delta r^2+z_c^2\right]^{3/2}}I_{eq}(\bar q\rightarrow 0,\bar\omega)\\
\nonumber
&=&\frac{S_0}{4}\frac{T^2}{T_c^2}\frac{\xi_0^2}{\varepsilon}\frac{z_c}{\left[\Delta r^2+z_c^2\right]^{3/2}}\\
&\times&\left\{\renewcommand{\arraystretch}{2.1} 
\begin{array}{cc} 
\displaystyle{1-\frac{2}{3}\left(\frac{\omega}{\omega_0}\right)^2} & \omega\ll \omega_0(T) \\
\displaystyle{\frac{\pi}{2}\frac{\omega_0}{\omega}} & \omega\gg\omega_0(T)
\end{array}\right. 
\renewcommand{\arraystretch}{1} 
.
\end{eqnarray}
As anticipated, the far field result, Eq. (\ref{eq:Seqlz}), exhibits asymptotic power-law behavior in both frequency 
and NV center positions, alongside a non-monotonic dependence on $z_c$ for large $\Delta r$. These characteristics are 
in agreement with the numerical results in Fig. \ref{fig:S2drz} and rely on the transverse current correlations 
$\langle J_\perp(\bq)J_\perp(-\bq)\rangle$ approaching a constant at small $q$. Such a behavior is consistent with a model 
in which both the squared amplitude of $\psi$ and its phase gradient are random fields with short-range correlations in 
space and time.

%\begin{figure*}[t!!!]
%\includegraphics[width=\textwidth,clip=true]{diagrams2nd.pdf}
%\vspace{10pt}
%\caption{\centering Second order diagrams for the current-current correlator.}
%\label{fig:diag2nd}
%\end{figure*}

\subsection{Non-equilibrium noise}

Next, we consider the system subject to an electric potential $\Phi(\br)=-(E/\qf) \sin (\bq_\phi\cdot\br)$, 
with $\bq_\phi=(\qf,0)$. At the end of the calculation we take the limit $\qf\rightarrow 0$, which results 
in a uniform static electric field ${\bm E}=E\hat{\bm x}$. Diagrammatically, the electric 
potential introduces new vertices into the theory, shown in Fig. \ref{subfig:vertices}. Using them to 
evaluate the average current density results in $\langle \bJ\rangle=\sigma_{AL} {\bm E}$, 
with the well-known AL contribution to the conductivity 
\begin{equation}
\sigma_{AL} = \frac{e^2}{16}\frac{T}{T_c}\frac{1}{\varepsilon}.
\end{equation}
The induced current generates a static magnetic field in the $y$ direction.   
Our focus, however, is the influence of the external electric field on the connected current correlations, 
$C_\perp(\bq,\nu)= \langle\langle J_\perp(\bq,\nu)J_\perp(-\bq,-\nu)\rangle\rangle$
%=\langle J_\perp(\bq,\nu)J_\perp(-\bq,-\nu)\rangle-\langle J_\perp(\bq,\nu)\rangle\langle J_\perp(-\bq,-\nu)\rangle$, 
which determine the magnetic noise. 

%\begin{figure}[t!!!]
%\includegraphics[width=\linewidth,clip=true]{diagrams2nd-onec.pdf}
%\vspace{10pt}
%\caption{\centering Second order diagrams for the current-current correlator.}
%\label{fig:diag2nd}
%\end{figure}

We emphasize that the electric field drives the superconducting fluctuations out of equilibrium 
without altering the thermal Langevin bath used in the TDGL description. In particular, the stochastic 
force $\zeta$ retains the equilibrium correlation function given in Eq. (\ref{eq:zetacorr}), with the bath maintained at temperature $T$. The electric field instead enters through the deterministic part of the 
TDGL equation and changes the steady-state distribution of the order parameter. Thus, while the 
equilibrium noise derived above satisfies the fluctuation-dissipation theorem, the resulting 
non-equilibrium current correlations are not constrained by it, (see also Ref. \cite{Romare26}). 

General conclusions regarding the behavior of $C_\perp$ can be derived from symmetry arguments. 
Consider the effect of 
charge conjugation, ${\cal C}:\psi(\br,t)\rightarrow\psi^*(\br,t)$, on the dynamics. If 
$\psi$ is a solution of the TDGL equation (\ref{eq:TDGL}), then $\psi^*$ satisfies the same equation 
with $-\tau'$, potential $-\Phi$ and noise $\zeta^*$. Since $J_\perp$ is odd under $\cal C$, namely 
$J_\perp(\psi)=-J_\perp(\psi^*)$, and because $\zeta^*$ and $\zeta$ share the same statistics, 
it follows that $C_\perp(E,\tau')=C_\perp(-E,-\tau')$. Hence, the $O(E)$ contribution to the noise 
is odd in $\tau'$ and therefore vanishes in the particle-hole symmetric case. Furthermore, 
in the Gaussian approximation $\tau'$ controls the reactive part of the TDGL dynamics. At linear 
order in $\tau'$, the correction contributes purely to the time-antisymmetric sector of the steady 
state correlator, and thus is odd in $\nu$.

Similar considerations apply to reflection about the $x$ axis, $\psi(x,y,t)\rightarrow \psi(x,-y,t)$. 
This symmetry dictates that $C_\perp(q,\phi,\nu)=C_\perp(q,-\phi,\nu)$, where we have used the parametrization 
$\bq=q(\cos\phi,\sin\phi)$. Thus, the angular dependence of the $O(E)$ contribution to $C_\perp$ 
is expected to be proportional to the only available scalar $\bm E\cdot\bq$.

%Regarding the frequency dependence of the current correlations we note that the uniform static field $E$ 
%advects the fluctuating order-parameter modes in momentum space, such that modes 
%contributing to the current at two times differ by $k_x(t)-k_x(t')=2eE(t-t')$. This by itself does not 
%create a handed asymmetry in the transverse current noise. 
%A non zero $\tau'$ adds chiral phase rotation to each mode, and the advection then makes the correlator 
%depend on the sign of $t-t'$, producing a decaying time-odd, and hence $\omega$-odd correction to $C_\perp$. 

\begin{figure}[t!!!]
\centering
\includegraphics[width=\columnwidth,clip=true]{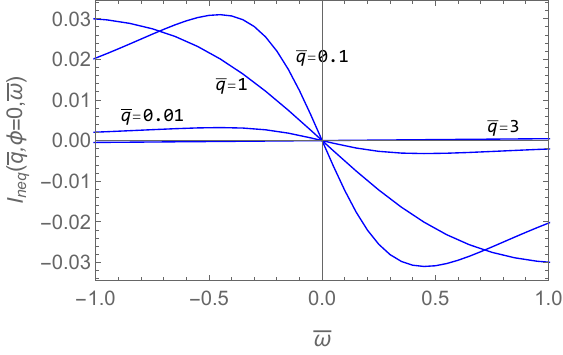}
\vspace{0pt}
\caption{$I_{neq}(\bar q,\phi=0,\bar\omega)$ for the case $\eta=0.1$, as a function of $\bar\omega$ for several values of $\bar q$.}
\label{fig:Ineq2d}
\end{figure}

The above features of the linear-in-$E$ contribution to the current fluctuations are confirmed by a direct calculation 
of the diagrams shown in Fig. \ref{subfig:1st}. Specifically, we present in Fig. \ref{fig:Ineq2d} numerical 
results for the frequency dependence of the $O(E)$ contribution 
\begin{eqnarray}
\nonumber
\hspace{-0.8cm}&&\langle\langle J_\perp(\bq,\nu)J_\perp(\bq',\nu')\rangle\rangle_{neq}=\\
\hspace{-0.8cm}&&-\frac{\sqrt{3}\pi^2f(E,\varepsilon) S_0}{\sqrt{2}}\frac{T^2}{T_c^2}\frac{\xi_0^2}{\varepsilon}
I_{neq}(\bar\bq,\bar\nu)\delta(\bq+\bq')\delta(\nu+\nu'), 
\end{eqnarray}
where we have introduced 
\begin{equation}
\label{eq:fdef}
f(E,\varepsilon)=\frac{\pi}{\sqrt{384}}\frac{eE\xi_0}{T_c}\frac{1}{\varepsilon^{3/2}},
\end{equation}
which provides a measure of the ratio between the non-equilibrium and equilibrium contributions 
to the noise, in a sense we make more precise below.

As evident from Fig. \ref{fig:Ineq2d}, the frequency dependence is approximately linear 
up to a frequency of order $\omega_0$. This allows us to obtain an analytical approximation 
for the kernel $I_{neq}$ in this range, which to $O(\eta)$ reads
\begin{eqnarray}
\label{eq:SE1}
\nonumber
\hspace{-0.5cm}&&I_{neq}(\bar\bq,\bar \nu)\equiv \eta I_{neq}(\bar q)\bar\nu\cos\phi=\\
\hspace{-0.5cm}&&\eta\frac{1}{\bar q(1+\bar q^2)^3}\left[11-\frac{11+6\bar q^2}{\bar q\sqrt{1+\bar q^2}}
{\rm arcsinh}(\bar q)\right]\bar\nu\cos\phi.
\end{eqnarray}

\begin{figure}[t!!!]
\centering
\includegraphics[width=\columnwidth,clip=true]{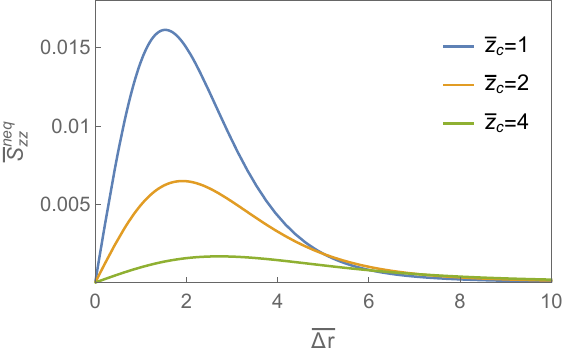}
\vspace{3pt}
\caption{$\bar S^{neq}_{zz}\equiv S^{neq}_{zz}/[i\omega\eta f(E,\varepsilon)(T/T_c)^2S_0/\omega_0(T)]$ as a function 
of $\overline{\Delta r}=\sqrt{\varepsilon}\Delta r/\xi_0$ for $\theta=0$ and several values of $\bar z_c =\sqrt{\varepsilon}z_c/\xi_0$.}
\label{fig:Sneq2d}
\end{figure}

Such current fluctuations contribute to the magnetic noise according to 
\begin{eqnarray}
\nonumber
S_{zz}^{neq}(\vr_1,\vr_2,\omega)&=&\eta\frac{\sqrt{3}f(E,\varepsilon) S_0}{\sqrt{2}}\frac{T^2}{T_c^2}\frac{i\omega}{\omega_0(T)}\cos\theta\\
\nonumber
&\times&\int_0^\infty d\bar q\, \bar q e^{-2\bar q z_c/\xi}J_1\left(\frac{2\bar q\Delta r}{\xi}\right)I_{neq}(\bar q),\\
\end{eqnarray}
where $\bm{\Delta r}=\Delta r(\cos\theta,\sin\theta)$, see Fig. \ref{fig:geometry2d}. 
Representative examples of the dependence on $\Delta r$ are shown in Fig. \ref{fig:Sneq2d}. 
An explicit expression can be derived in the case $z_c\gg\xi_0/\sqrt{\varepsilon}$, where the integral is dominated 
by small $\bar q$, for which $I_{neq}=4\bar q/3$. As a result we find in this regime 
\begin{eqnarray}
\label{eq:E1signal}
\nonumber
S_{zz}^{neq}(\vr_1,\vr_2,\omega)&=&\eta\frac{\sqrt{3}\pi f(E,\varepsilon) S_0}{2^{13/2}}\frac{T^2}{T_c^3}\frac{\xi_0^3}{\varepsilon^{5/2}}\\
&\times&i\omega\frac{\Delta r\, z_c}{(\Delta r^2+z_c^2)^{5/2}}\cos\theta.
\end{eqnarray}

Owing to the typical smallness of the symmetry breaking parameter $\eta=\tau'/\tau$, nonlinear effects 
may dominate the non-equilibrium noise. Hence, we proceed to calculate the $O(E^2)$ contribution in the 
particle-hole symmetric case $\eta=0$. Concentrating on the dc limit, we obtain for the connected transverse 
current-current correlations 
\begin{eqnarray}
\nonumber
\hspace{-0.8cm}&&\lim_{\nu\rightarrow 0}\langle\langle J_\perp(\bq,\nu)J_\perp(\bq',\nu')\rangle\rangle_{neq-ph}=\\
\hspace{-0.8cm}&&\;\;\;\;\;\;-\frac{3\pi^2 S_0}{16}\frac{T^2}{T_c^2}\frac{\xi_0^2}{\varepsilon}
f^2(E,\varepsilon) I_{neq}^{ph}(\bar\bq)\delta(\bq+\bq')\delta(\nu+\nu'), 
\end{eqnarray}
where $ph$ signifies the fact that this is the leading term due to weak $E$ in the case $\eta=0$, and 
\begin{equation}
I_{neq}^{ph}(\bar\bq)=I_{neq,0}^{ph}(\bar q)+I_{neq,2}^{ph}(\bar q)\cos 2\phi,
\end{equation}
is expressed in the parameterization $\bar\bq=\bar q(\cos\phi,\sin\phi)$ via
\begin{eqnarray}
\nonumber
I_{neq,0}^{ph}(\bar q)&=&-\frac{1}{\bar q^2(1+\bar q^2)^3}\Bigg [11-10\bar q^2-\frac{8\bar q^4}{3}\\
&&-\frac{11-8\bar q^2}{\bar q\sqrt{1+\bar q^2}}{\rm arctanh}\left(\frac{\bar q}{\sqrt{1+\bar q^2}}\right)\Bigg], \\
\nonumber
I_{neq,2}^{ph}(\bar q)&=& -\frac{1}{\bar q^2(1+\bar q^2)^3}\Bigg[ -1+4\bar q^2 +4\bar q^4 \\
&&+\frac{1+2\bar q^2}{\bar q\sqrt{1+\bar q^2}}{\rm arctanh}\left(\frac{\bar q}{\sqrt{1+\bar q^2}}\right)\Bigg]. 
\end{eqnarray}

The above result arises because, at order $E^2$, the fluctuations are biased along the drive direction, 
thereby imparting a nematic character to their ensemble. Specifically, it reflects the fact that the 
$\langle J_y J_y\rangle$ correlations acquire a negative constant contribution in the $q\rightarrow 0$ 
limit, while the other components receive only $O(q^2)$ corrections. This also leads to the quadrupole 
structure of the transverse current correlations in the same limit. 

\begin{figure}[t!!!]
\centering
\includegraphics[width=\columnwidth,clip=true]{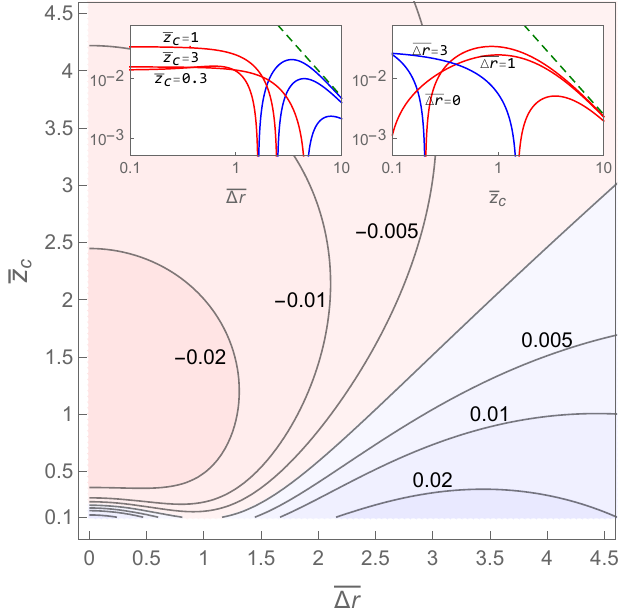}
\vspace{10pt}
\caption{Contour plot of $S_{zz}^{neq-ph}(\omega=0)/[S_0f^2(E,\varepsilon)(T/T_c)^2]$ 
as a function of $\overline{\Delta r}=\sqrt{\varepsilon}\Delta r/\xi_0$ and $\bar z_c 
=\sqrt{\varepsilon}z_c/\xi_0$ for $\theta=0$. The left inset shows several $\overline{\Delta r}$ cuts, 
with the dashed line indicating a $\overline{\Delta r}^{\,-2}$ decay. 
The right inset depicts $\bar z_c$ cuts, where the dashed line represents a $\bar z_c^{\,-2}$ decay. 
Red (Blue) segments correspond to negative (positive) values.}
\label{fig:S2dneq-ph}
\end{figure}

\begin{figure}[t!!!]
\centering
\includegraphics[width=\columnwidth,clip=true]{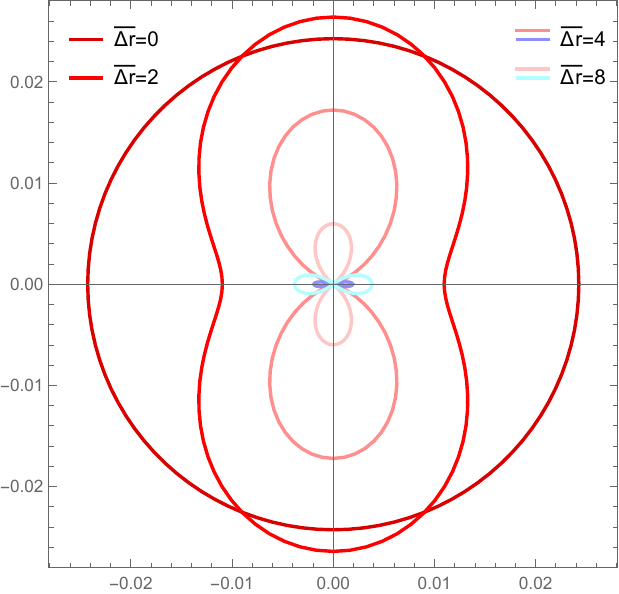}
\vspace{10pt}
\caption{Polar representation of $S^{neq-ph}_{zz}/[S_0f^2(E,\varepsilon)(T/T_c)^2]$ for the case $z_c=2\xi_0/\sqrt{\varepsilon}$ 
and various values of $\overline{\Delta r}=\Delta r\sqrt{\varepsilon}/\xi_0$. Red (Blue) segments correspond to negative (positive) values.}
\label{fig:angdep2d}
\end{figure}

Using the form of $I_{neq}^{ph}(\bar\bq)$, the $O(E^2)$ non-equilibrium contribution to the magnetic noise spectrum becomes
\begin{eqnarray}
\label{eq:S2dneq}
\nonumber
&&S^{neq-ph}_{zz}(\vr_1,\vr_2,0) \\
\nonumber
&&=\frac{3S_0}{32\pi}\frac{T^2}{T_c^2}f^2(E,\varepsilon) 
\int d^2\bar q \, e^{-2(\bar q z_c-i\bar \bq\cdot {\bm {\Delta r}})/\xi} I_{neq}^{ph}(\bar \bq) \\
\nonumber
&&=\frac{3S_0}{16}\frac{T^2}{T_c^2}f^2(E,\varepsilon)\int_0^\infty d\bar q\, \bar q e^{-2\bar q z_c/\xi} \\
\nonumber
&&\times \left[ J_0\!\left(\frac{2\bar q\Delta r}{\xi}\right)I_{neq,0}^{ph}(\bar q) 
-J_2\!\left(\frac{2\bar q\Delta r}{\xi}\right)I_{neq,2}^{ph}(\bar q)\cos 2\theta\right],\\
\end{eqnarray}
where we continue to use $\bm{\Delta r}=\Delta r(\cos\theta,\sin\theta)$ and $J_{0,2}$ are Bessel functions.
The dependence of this contribution on the spatial configuration of the sensors   
is illustrated in Figs. \ref{fig:S2dneq-ph} and \ref{fig:angdep2d}. In the regime where $z_c>\xi$ and $z_c\gg\Delta r$, 
it is isotropic, negative and decays as $z_c^{-2}$. Equation (\ref{eq:S2dneq}) shows that $z_c$ acts as a lateral 
coarse-graining scale, effectively filtering out fluctuation modes with $q > 1/z_c$. Hence,  
when $z_c\gg\Delta r$, the factor $\exp(i\bq \cdot\Delta_r)\approx1$ and the two probes sample  
the same blurred field patch. The amplitude of the resulting isotropic signal is proportional to the number 
of contributing modes, i.e. $z_c^{-2}$, and the elongation of current fluctuations by the drive suppresses the 
generated magnetic field, resulting in a negative correction to the equilibrium result. Conversely, when 
$\Delta r>\xi$ and $\Delta r\gg z_c$, the signal follows a $\cos2\theta$ angular dependence and decays as $\Delta r^{-2}$.
Under such conditions the sensor separation $\Delta r$ serves as the dominant filtering scale in Eq. (\ref{eq:S2dneq}),  
suppressing modes with $q > 1/\Delta r$ and resulting in the $\Delta r^{-2}$ scaling. 
Unlike the previous case, the sensors here are sufficiently close to the plane to resolve the quadrupolar distortion 
of the transverse current correlations induced by the driving field.

An analytical expression for the magnetic noise, which encapsulates the aforementioned features, can be derived 
in the limit $z_c \gg \xi_0/\sqrt{\varepsilon}$ by employing the small-$\bar{q}$ asymptotic form of the transverse 
current kernel
\begin{equation}
\label{eq:Iphneqsmallq}
I_{neq}^{ph}(\bq)=-\frac{16}{3}\left(1+\cos2\phi\right)+\frac{64}{15}\left(7+3\cos2\phi\right)\bar q^2.
\end{equation}
As a result
\begin{eqnarray}
\label{eq:E2signal}
\nonumber
&&\!\!\!S^{neq-ph}_{zz}(\vr_1,\vr_2,0)\\
\nonumber
%&&=-\frac{\pi^3 e^4E^2\xi_0^2T^2}{1536T_c^3\varepsilon^4}
&&\!\!\!=-\frac{S_0}{4}\frac{T^2}{T_c^2}\frac{\xi_0^2}{\varepsilon}f^2(E,\varepsilon)\\
\nonumber
&&\!\!\!\;\;\;\;\times
\Bigg\{\frac{z_c}{(\Delta r^2 +z_c^2)^{3/2}}
+\frac{21\xi_0^2 z_c(3\Delta r^2-2z_c^2)}{5\varepsilon(\Delta r^2+z_c^2)^{7/2}}-\frac{1}{\Delta r^2}\\
\nonumber
&&\!\!\!\;\;\;\;\times\left[2-\frac{z_c(3\Delta r^2+2z_c^2)}{(\Delta r^2+z_c^2)^{3/2}}-\frac{9\xi_0^2 z_c\Delta r^4}
{\varepsilon(\Delta r^2+z_c^2)^{7/2}}\right]\cos2\theta\!\Bigg\}\\
\nonumber
&&\!\!\!=\frac{S_0}{4}\frac{T^2}{T_c^2}\frac{\xi_0^2}{\varepsilon}f^2(E,\varepsilon)\\
&&\!\!\!\;\;\;\;\times\left\{\renewcommand{\arraystretch}{2.5} 
\begin{array}{cc} 
\displaystyle{-\frac{1}{z_c^2}+\frac{42\xi_0^2}{5\varepsilon z_c^4}+ \frac{3\Delta r^2}{4z_c^4}\left(2+\cos2\theta\right)} & \Delta r\ll z_c \\
\displaystyle{\frac{2}{\Delta r^2}\cos2\theta-\frac{z_c}{\Delta r^3}\left(1+3\cos2\theta\right)} & 
\Delta r \gg z_c
\end{array}\right. 
\renewcommand{\arraystretch}{1} 
\!\!\!\!.
\end{eqnarray}
Comparing this with Eq. (\ref{eq:Seqlz}) indicates that $f^2(E,\varepsilon)$ provides a measure 
of the ratio between the non-equilibrium and equilibrium contributions to the far field, 
$z_c\gg {\rm max}(\xi_0/\sqrt{\varepsilon},\Delta r)$, noise.

\section{Comparison with vortex fluctuations}
\label{sec:vortexfluct}

It is useful to contrast the Gaussian results with the magnetic noise expected from a vortex liquid. 
The equilibrium vortex problem has been analyzed in detail 
in Ref. \cite{Curtis24}, including both free vortices and bound vortex-antivortex pairs. 
Here, we also consider the non-equilibrium response of free vortices above $T_{\rm BKT}$,  
and its signatures in nonlocal noise measurements. 

To this end, we coarse grain on a scale $a_c$, integrating out tightly bound vortex-antivortex 
pairs with smaller separations, and describe the remaining vortices as a Coulomb gas of charges
$q_i=\pm1$ at positions $\bR_i(t)$. 
Each vortex follows an overdamped Langevin dynamics \cite{AHNS,Minnhagen,Dorsey-vortex}, which in the 
presence of a transport sheet current $\bJ$ reads
\begin{equation}
\label{eq:vor-Langevin}
(\alpha -\alpha_Hq_i\hat z\times)\dot\bR_i
=q_i\left[-\bm\nabla\varphi(\bR_i)-\phi_0\hat z\times \bJ\right]+\bm\eta_i.
\end{equation}
Here, $\phi_0=\pi/e$ is the flux quantum, and the potential induced by the vortices
\begin{equation}
\label{eq:vortex-potential}
\varphi(\br,t)=-2\pi\bar\rho_s\int d^2r'\ln\frac{|\br-\br'|}{a_c}\sigma(\br',t),
\end{equation}
is expressed via the vorticity density
\begin{equation}
\label{eq:vorticity-density}
\sigma(\br,t)=\sum_i q_i\delta[\br-\bR_i(t)],
\end{equation}
with $\bar\rho_s\equiv\rho_s(a_c)$ the effective phase stiffness renormalized on scales below $a_c$. 
It should not be confused with the long-distance stiffness, which vanishes above $T_{\rm BKT}$.
Finally, the Gaussian force noise satisfies 
\begin{equation}
\label{eq:vortex-noise}
\langle\eta_i^\mu(t)\eta_j^\nu(t')\rangle=2\alpha T\delta_{ij}\delta_{\mu\nu}\delta(t-t').
\end{equation}
The appearance of both longitudinal and Hall viscous drag coefficients, $\alpha$ and $\alpha_H$, 
in Eq. (\ref{eq:vor-Langevin}) was demonstrated in a TDGL calculation by Dorsey \cite{Dorsey-vortex}. 
Specifically, he showed that for vanishing normal Hall conductivity and small imaginary TDGL 
relaxation rate $\tau'$ they scale as $\alpha=\alpha_0+O(\tau'^2)$ and 
$\alpha_H=\alpha_{H,1}\tau'+O(\tau'^3)$.

Next, we define the vortex mobilities $\mu=\alpha/(\alpha^2+\alpha_H^2)$, 
$\mu_H=\alpha_H/(\alpha^2+\alpha_H^2)$ and invert Eq. (\ref{eq:vor-Langevin}) to obtain 
\begin{equation}
\label{eq:vort-dyn-invert}
\dot\bR_i=-(\mu q_i+\mu_H\hat z\times)\bm\nabla\varphi(\bR_i)+q_i\bu_E+\bw_E+\bm \xi_i,
\end{equation}
where we have decomposed the drive-induced velocity into a vorticity-dependent transverse motion and 
a common drift of all vortices 
\begin{equation}
\label{eq:vel-decomp}
\bu_E=\phi_0\mu\bJ\times\hat z, \;\;\;\;\;\;\;\;\;\;\; \bw_E=\phi_0\mu_H\bJ.
\end{equation}
Defining $D=\mu T$, the noise $\bm\xi_i=(\mu+q_i\mu_H\hat z\times)\bm \eta_i$ satisfies
\begin{equation}
\label{eq:xi-noise}
\langle\xi_i^\mu(t)\xi_j^\nu(t')\rangle=2D\delta_{ij}\delta_{\mu\nu}\delta(t-t'). 
\end{equation}
Starting from Eq. (\ref{eq:vort-dyn-invert}), one can use Ito calculus to derive a stochastic continuity 
equation \cite{Dean} $\partial_t n_q +\bm\nabla\cdot \bm j_q=0$, for the coarse-grained densities of free 
vortices $n_q(\br,t)=\sum_{i\in q}\delta[\br-\bR_i(t)]$, with 
\begin{equation}
\bm j_q=-D\bm\nabla n_q-(\mu q +\mu_H\hat z \times)n_q\bm\nabla\varphi+(q\bu_E+\bw_E)n_q+\bm\zeta_q,
\end{equation}
and $\bm \zeta_q(\br,t)=\sum_{i\in q}\delta[\br-\bR_i(t)]\bm\xi_i(t)$ obeying
\begin{equation}
\langle\zeta_q^\mu(\br,t)\zeta_{q'}^\nu(\br',t')\rangle=2Dn_q(\br,t)\delta_{qq'}\delta_{\mu\nu}\delta(\br-\br')\delta(t-t').
\end{equation}
Adding and subtracting the equations for the two species result in coupled dynamics 
for $\sigma(\br,t)$ and the total vortex density $\rho(\br,t)=\sum_i\delta[\br-\bR_i(t)]$. 
Upon linearizing about the average homogeneous vortex density $\rho(\br,t)=\rho_0+\delta\rho(\br,t)$  
the dynamics takes the form 
\begin{eqnarray}
\label{eq:stochsig} 
\nonumber
    &&\hspace{-0.7cm}\left[\partial_t -D\left({\bm \nabla}^2-\kappa^2\right)+\bw_E\cdot\bm\nabla \right]\sigma
    +\bm u_E\cdot\bm \nabla \delta\rho= \\
    &&\hspace{-0.7cm}\hspace{5.7cm}-\bm\nabla\cdot\bm\zeta_\sigma,\\
\label{eq:stochdrho}
\nonumber
   && \hspace{-0.7cm}\left[\partial_t-D{\bm \nabla}^2+\gamma_\rho+\bw_E\cdot\bm\nabla\right]\delta\rho
   +\bm u_E\cdot\bm \nabla \sigma =\\
   &&\hspace{-0.7cm}\hspace{5.7cm}-\bm\nabla\cdot\bm\zeta_\rho+\zeta_R.
\end{eqnarray}
Here,
\begin{equation}
\label{eq:kappa2}
\kappa^2=\frac{4\pi^2\bar\rho_s\rho_0}{T},
\end{equation}
and within the linear approximation $\bm\zeta_{\sigma,\rho}=\bm\zeta_+\pm\bm\zeta_-$ follow 
\begin{equation}
\label{eq:zetasigrho}
\langle\zeta_c^\mu(\br,t)\zeta_{c'}^\nu(\br',t')\rangle=2D\rho_0\delta_{cc'}
\delta_{\mu\nu}\delta(\br-\br')\delta(t-t'),
\end{equation}
where $c=\sigma,\rho$. 
Finally, when deriving Eqs. (\ref{eq:stochsig}) and (\ref{eq:stochdrho}) we took notice of 
the fact that the above continuity equations for $n_\pm$ do not contain conversion between 
short-range vortex-antivortex pairs and free vortices. 
Such pair dissociation and recombination events are standard ingredients of vortex dynamics 
\cite{AHNS,Minnhagen}. They relax the total free-vortex density but leave the net vorticity 
unchanged. Consequently, we have included their effects through 
a relaxation term $-\gamma_\rho\delta\rho$ and a corresponding reaction noise $\zeta_R$, 
chosen so 
\begin{equation}
\label{eq:zetaR}
\langle\zeta_R(\br,t)\zeta_R(\br',t')\rangle=2\gamma_\rho\rho_0\delta(\br-\br')\delta(t-t').
\end{equation}
We note that this choice, together with Eq. (\ref{eq:zetasigrho}), leads via Eq. (\ref{eq:stochdrho})  
to an equilibrium structure factor 
$\langle\delta\rho(\br,t)\delta\rho(\br',t)\rangle=\rho_0\delta(\br-\br')$, consistent with the 
dilute free-vortex approximation adopted here. 

The vorticity fluctuations contribute directly to the transverse supercurrent fluctuations via 
\begin{equation}
\label{eq:vortexJperp}
j_\perp^v(\bk,t)=i\frac{4\pi e\bar\rho_s}{k}\sigma(\bk,t).
\end{equation}
Thus, using 
$\langle\sigma(\bk,\omega)\sigma(\bk',\omega')\rangle=C_{\sigma\sigma}(\bk,\omega)(2\pi)^3\delta(\bk+\bk')\delta(\omega+\omega')$, the vortex contribution to the 
nonlocal magnetic-noise spectrum takes form 
\begin{equation}
\label{eq:Szzv}
S_{zz}^v(\bm{\Delta r},z_c,\omega)=(4\pi e\bar\rho_s)^2\int d^2k\, e^{i\bk\cdot\bm{\Delta r}-k z_c}
\frac{C_{\sigma\sigma}(\bk,\omega)}{k^2}.
\end{equation}
%This expression neglects magnetic screening and therefore applies on scales shorter than 
%the Pearl length.
In equilibrium $\bu_E=\bw_E=0$ and Eq. (\ref{eq:stochsig}) yields
\begin{equation}
\label{eq:Csigeq}
C_{\sigma\sigma}(\bk,\omega)=\frac{2D\rho_0 k^2}{\omega^2+D^2(k^2+\kappa^2)^2},
\end{equation}
leading, in the limit $\kappa z_c\gg1$, to the equilibrium signal
\begin{equation}
\label{eq:Sveqlargez}
S_{zz}^{eq,v}(\Delta r,z_c,\omega)=S_0^v\frac{T}{T_{\rm BKT}}\frac{1}{1+(\omega/\omega_v)^2}
\frac{z_c}{\kappa^2(\Delta r^2+z_c^2)^{3/2}}.
\end{equation}
Here, we have introduced the typical noise scale 
\begin{equation}
\label{eq:S0v}
S_0^v=\frac{16\pi e^2 T_{\rm BKT}\bar\rho_s}{D},
\end{equation}
and the characteristic frequency 
\begin{equation}
\label{eq:omegav}
\omega_v=D\kappa^2.
\end{equation}
Comparing Eq. (\ref{eq:Sveqlargez}) with the corresponding Gaussian result, Eq. (\ref{eq:Seqlz}), 
reveals that both exhibit a low-frequency plateau, followed by an $\omega^{-p}$ decay, with $p=1,2$ 
for Gaussian and vortex fluctuations, respectively. Another key distinction is their crossover 
frequency, where the decay commences. While the Gaussian crossover is dictated by the Cooper-pair 
inverse lifetime $\omega_0$ (typically in the THz regime near $T_c$), the characteristic BKT frequency 
$\omega_v$ is orders of magnitude smaller, see Sec. \ref{sec:exp}. Furthermore, the full equilibrium 
theory of Ref. \cite{Curtis24} includes a second plateau at higher frequencies due to surviving 
bound vortex-antivortex pairs. Both the Gaussian and vortex models share a similar far-field spatial 
dependence. However, this dependence emerges once $z_c$ crosses different characteristic scales: $\xi(T)$  
and the BKT correlation length $\xi_+(T)\sim\kappa^{-1}$, respectively. Their distinct 
temperature dependence allows one to distinguish between the two scenarios. 

For $\kappa z_c\ll1$ and $\omega=0$ we obtain 
\begin{equation}
\label{eq:Sveqsmallz}
S_{zz}^{eq,v}(\Delta r,z_c,0)=S_0^v\frac{T}{T_{\rm BKT}}
\left\{\renewcommand{\arraystretch}{2.1} 
\begin{array}{cc} 
 \displaystyle{\frac{1}{4}(2-\pi\kappa z_c)} & 
\displaystyle{\;\;\Delta r\ll\kappa^{-1}} \\
\displaystyle{\frac{1}{\kappa^2}\frac{z_c}{\Delta r^3}} & \displaystyle{\;\;\Delta r\gg \kappa^{-1}}
\end{array}\right.\!\!, 
\end{equation}
to be compared with Eq. (\ref{eq:Seqdr0omega0}). 

Turning to the driven system, we note that $\bJ\propto \bm E$ for sufficiently weak $E$ above $T_{\rm BKT}$. 
Thus Eq. (\ref{eq:vel-decomp}) implies $u_E=O(E)$ and $w_E=O(\tau'E)$. The invariance of the TDGL 
equation under charge conjugation $\psi\rightarrow\psi^*$, $\tau'\rightarrow -\tau'$, $E\rightarrow-E$, 
$\zeta\rightarrow\zeta^*$ is then mirrored by the invariance of the vortex dynamics, Eqs. (\ref{eq:stochsig}) 
and (\ref{eq:stochdrho}), under $\sigma\rightarrow-\sigma$, $u_E\rightarrow-u_E$, 
$\zeta_\sigma\rightarrow -\zeta_\sigma$, with the result that 
$C_{\sigma\sigma}(u_E,w_E)=C_{\sigma\sigma}(-u_E,w_E)$. 
Consequently, a nonzero $O(E)$ response requires nonzero $w_E$, i.e., broken particle-hole symmetry. 
Indeed, at $O(E)$ we find 
\begin{equation}
\label{eq:CsigO1}
\delta C_{\sigma\sigma}^{(1)}(\bk,\omega)=\frac{4D\rho_0(\bw_E\cdot\bk)\omega k^2}{\left[\omega^2+D^2(k^2+\kappa^2)^2\right]^2},
\end{equation}
which in the far-field $\kappa z_c\gg1$ and for $\omega\ll\omega_v$ yields, 
\begin{eqnarray}
\label{eq:Sv1largez}
\nonumber
\delta S_{zz}^{(1),v}(\bm{\Delta r},z_c,\omega)&=&6S_0^v\frac{T}{T_{\rm BKT}}\frac{\kappa w_E}{\omega_v} \\
&\times&
\frac{i\omega}{\omega_v}\frac{1}{\kappa^3}\frac{\Delta r\, z_c}{(\Delta r^2+z_c^2)^{5/2}}
\cos\theta.
\end{eqnarray}
The structure of Eq. (\ref{eq:Sv1largez}) closely parallels that of the Gaussian $O(E)$ result, 
Eq. (\ref{eq:E1signal}), albeit with a scaling that is governed by $\xi_+$ and $\omega_v$ instead 
of $\xi$ and $\omega_0$.

Specializing to the static particle-hole symmetric case $w_E=\omega=0$, we find at $O(E^2)$
\begin{equation}
\label{eq:CsigO2}
\delta C_{\sigma\sigma}^{(2,ph)}(\bk,0)=\frac{2\rho_0(\bu_E\cdot\bk)^2(\kappa^2-k^2)}{D^3(k^2+\kappa^2)^3(k^2+\ell_\rho^{-2})},
\end{equation}
where 
\begin{equation}
\label{eq:lrhodef}
\ell_\rho=\sqrt{D/\gamma_\rho}
\end{equation}
is the distance over which a fluctuation of the total free-vortex density typically diffuses 
before being relaxed by vortex-antivortex dissociation and recombination.
%is the distance over which a vortex can diffuse before its identity as part of a total-density 
%fluctuation is lost by pair reactions. 
This leads, for $z_c\gg\kappa^{-1},\ell_\rho$, to 
\begin{eqnarray}
\label{eq:eq:Sv2largez}
\nonumber
\!\!\!\!\!\!\!\!S_{zz}^{(2),v}(\Delta r,z_c,0)&=&\frac{S_0}{2}^v\frac{T}{T_{\rm BKT}}
\left(\frac{u_E\kappa}{\omega_v}\right)^2 \ell_\rho^2\\
\!\!\!\!\!\!\!\!&\times&
\left\{\renewcommand{\arraystretch}{2.1} 
\begin{array}{cc} 
 \displaystyle{\frac{1}{z_c^2}} & 
\displaystyle{\;\;\Delta r\ll z_c} \\
\displaystyle{\frac{2}{\Delta r^2}\cos 2\theta} & \displaystyle{\;\;\Delta r\gg z_c}
\end{array}\right.\!\!.
\end{eqnarray}
Notably, for $\Delta r\ll z_c$ it contains a positive isotropic contribution, in contrast to the 
negative contribution which appears in the Gaussian result, Eq. (\ref{eq:E2signal}). 
This increase can be traced back to the $\kappa^2$ piece in the numerator of Eq. (\ref{eq:CsigO2}), 
which in turn originates from the square of first order $u_E$ processes in Eq. (\ref{eq:stochsig}). 
Physically, it occurs because the applied field drives vortices and antivortices 
in opposite directions, thereby converting fluctuations of the total free-vortex density into 
additional vorticity fluctuations. Conversely, no analogous density-to-vorticity conversion 
channel exists in the Gaussian scenario. Instead, the field distorts the fluctuating Cooper-pair 
spectrum itself, leading to a reduction of the long-wavelength isotropic 
transverse-current fluctuations.

In the near-field, $z_c\ll\kappa^{-1}$, and taking for illustration $\ell_\rho=\kappa^{-1}$, we find
\begin{eqnarray}
\label{eq:Sv2smallz}
\nonumber
&&S_{zz}^{(2),v}(\Delta r,z_c,0)=\frac{S_0}{2}^v\frac{T}{T_{\rm BKT}}
\left(\frac{u_E\kappa}{\omega_v}\right)^2 \\
\nonumber
&&\times
\left\{\renewcommand{\arraystretch}{2.1} 
\begin{array}{cc} 
 \displaystyle{\frac{1}{12}-\frac{\kappa^2 (2z_c^2-\Delta r^2)}{48}-\frac{\kappa^2\Delta r^2}{96}\cos 2\theta} & 
\displaystyle{\;\;\Delta r\ll \kappa^{-1}} \\
\displaystyle{\frac{2\ell_\rho^2}{\Delta r^2}\cos 2\theta} & \displaystyle{\;\; \Delta r\gg \ell_\rho}
\end{array}\right.\!\!. \\
\end{eqnarray}
Here, for $z_c,\Delta r\ll\kappa^{-1}$ where larger wavevectors come into play, the $-k^2$ piece in the 
numerator of Eq. (\ref{eq:CsigO2}) plays an important role. It originates from a product of the $u_E$ 
terms in Eqs. (\ref{eq:stochsig}) and (\ref{eq:stochdrho}), and corresponds to a process where an 
existing vorticity fluctuation is converted by the field into an elongated density fluctuation along 
the $y$ direction, which in turn feeds back into the vorticity. This acts as an additional relaxation 
channel for the original vorticity modulation that reduces the isotropic signal, although the $\kappa^2$ 
term still wins. More importantly, it produces a $-\cos 2\theta$ anisotropy with a positive response 
in the direction perpendicular to the current. This feature that distinguishes it from the other cases 
actually persists for general $\kappa^{-1}/\ell_{\rho}$. 

The discussion above treats the mean density of free vortices, $\rho_0$, as fixed. A sufficiently 
strong current can, however, also modify $\rho_0$ itself. The Magnus force exerted by the current 
tilts the logarithmic vortex-antivortex interaction potential and produces a saddle at a current-dependent 
separation $r_c\simeq2\pi\rho_s(r_c)/\phi_0 J$. Thermally activated diffusion of a pair beyond this point  produces additional free vortices. 
A recent study, Ref. \cite{Sarang-vortices26}, considered their contribution to the 
magnetic noise measured by a pair of NV centers slightly below $T_{\rm BKT}$, and found an enhanced 
signal when the pair is aligned perpendicular to the current. Above $T_{\rm BKT}$, however, 
an equilibrium density of free vortices is already present and screens the logarithmic interaction 
beyond the BKT correlation length $\xi_+\sim\kappa^{-1}$. Consequently, current-induced unbinding 
becomes important only when $r_c$ drops below $\xi_+$. This defines an onset current 
\begin{equation}
\label{eq:J*}
J_*=\frac{2\pi\rho_s(\xi_+)}{\phi_0\xi_+}\simeq\frac{4T}{\phi_0\xi_+},
\end{equation}
where we have used the fact that close to the BKT separatrix $\rho_s(\xi_+)\simeq2T/\pi$. 
%Since the linear vortex-flow resistance above $T_{\rm BKT}$ is $R^v=\mu\phi_0^2\rho_0$, this 
%translates into an electric field 
%\begin{equation}
%\label{eq:E*}
%E_*=\mu\phi_0^2\rho_0 J_*.
%\end{equation}

For $J\ll J_*$, screening occurs before the current can tear apart pairs, and the 
mean free-vortex density changes only perturbatively. By analyticity the change is expected to 
be quadratic in the field   
$\rho(E)=\rho_0+\delta\rho^{(2)}(E)$. This adds to the noise a term 
$(\partial S_{zz}^{eq,v}/\partial\rho_0)\delta\rho^{(2)}(E)$. Since in the $\omega=0,\ k\ll\kappa$ 
limit $S_{zz}^{eq,v}\sim 1/\rho_0$, it produces an isotropic negative contribution 
$-(S_{zz}^{eq,v}/\rho_0)\delta\rho^{(2)}(E)$. Physically, adding free vortices increases 
the instantaneous number of magnetic sources, but it simultaneously shortens the screening length 
and for the prescribed conditions the latter effect wins.

For $J> J_*$, current-induced pair production becomes an essential part of the dynamics. 
As mentioned, Ref. \cite{Sarang-vortices26} finds that current-generated vortices below $T_{\rm BKT}$ 
produce enhanced correlations for probes separated perpendicular to the current. Above $T_{\rm BKT}$, 
we expect an analogous $-\cos2\theta$ contribution from current-induced vortices, 
through a mechanism similar to the one discussed below Eq. (\ref{eq:Sv2smallz}), provided 
the probes resolve distances shorter than the current-dependent screening length.
Because the anisotropic Gaussian component, Eq. (\ref{eq:E2signal}), is 
proportional to $\cos 2\theta$, this raises the interesting possibility of observing a sign reversal 
in the anisotropy. Given that the density of induced vortices scales as $\xi_+^{-2}(J/J_*)^2$ \cite{AHNS}, 
we expect the current induced component to become dominant and the sign reversal to occur at a current of 
order $J_*$. As the current increases, the additional free vortices shorten the current-dependent screening length $\kappa^{-1}(J)$. Once it becomes comparable to the larger of $z_c$ and $\Delta r$, the probes 
cease to resolve the unscreened drift regime, so the negative quadrupolar component should weaken and may eventually cross over to the long-wavelength positive anisotropy.

\begin{table*}[t!] 
    \centering 
    \setlength{\tabcolsep}{2.8 pt}
    \begin{tabular}{@{} l c c c c | c c @{}} 
        \toprule 
        Material & $T_c$ (K) & $\xi_{ab}(0)$ (nm) & $\xi_c(0)$ (nm) & c (nm) & $S_0/2\pi$ (pT$^2$/Hz) & $f^2/E^2$ (V/cm)$^{-2}$ \\
        \midrule 
        Bulk La$_{1.91}$Sr$_{0.09}$CuO$_4$ \cite{Kimura-LSCO} & 29 & 3 & 0.075 & 1.3 & 67 & $7\times 10^{-6}$ \\
        46-nm-thick  \BSCCOs film \cite{BSCCO-Aichner2025} & 90 & 1.8 & 0.01 & 3.1 & 577 & $8\times 10^{-6}$ \\
        4-nm-thick FeSe/SrTiO$_3$ \cite{FeSe-STO-Kobayashi}  & 30 & 2.7 & 0.18 & 0.55 & 85 & $6\times 10^{-6}$ \\
        26 atomic layer Pb film \cite{Pb-Guo}  & 6.5 & 83 (bulk) & 83 (bulk) & & 0.02 & $10^{-3}$ \\
        10 layer NbSe$_2$ film \cite{NbSe2-Wang} & 4.6 & 18 & 2.7 (bulk) & 1.25 & 0.3 & $4\times 10^{-5}$ \\
        \bottomrule 
    \end{tabular}
    \vspace{20pt}
    \caption{\label{tab:exp} Experimental parameters for representative quasi-two-dimensional superconducting systems. 
    $\xi_{ab}(0)$ and $\xi_c(0)$ denote the zero-temperature in-plane and out-of-plane coherence lengths, respectively, 
    while $c$ is the $c$-axis lattice constant. $S_0/2\pi$ sets the scale for the equilibrium magnetic noise. The 
    dimensionless ratio $f^2$ represents the $O(E^2)$ non-equilibrium contribution relative to the equilibrium 
    noise, evaluated in the far-field limit for the system at $T = T_c + 1$ K.}
\end{table*}

\section{Experimental considerations}
\label{sec:exp}

To assess the experimental detectability of the magnetic noise using current NV sensing 
platforms, Table \ref{tab:exp} provides the relevant physical parameters and the 
characteristic equilibrium noise scale $S_0/2\pi$ for several representative 
quasi-two-dimensional systems. As indicated by Eq. (\ref{eq:S0def}) and illustrated 
by the table, the equilibrium noise is significantly enhanced in high-temperature superconductors 
relative to conventional systems, driven by their higher $T_c$ and shorter in-plane coherence lengths. 
Given that the current detection limit for magnetic noise spectral densities via $T_1$ relaxometry 
is approximately $10~\text{pT}^2/\text{Hz}$ \cite{Andersen2019}, our analysis suggests that \BSCCOs 
(Bi2212) films are particularly favorable for experimental observation. 

The Gaussian approximation employed above eventually breaks down sufficiently close to $T_c$, 
where interactions between order-parameter fluctuations associated with the quartic term in 
Eq. (\ref{eq:GLF}) become important. The corresponding crossover is conventionally characterized 
by the Ginzburg number $Gi$, with the Gaussian treatment expected to become reliable 
for reduced temperatures $\varepsilon> Gi$. In highly anisotropic cuprates, and particularly 
in layered Bi2212, the precise extent of the critical region depends on the sample geometry. 
Recent analyses of fluctuation conductivity in ultrathin Bi2212 films, 
using data for films containing 4, 6, and 10 CuO$_2$ bilayers, obtained $Gi=0.009$, 0.005, 
and 0.004, respectively \cite{Botana22}. Moreover, low-field fluctuation-magnetoconductivity 
measurements on bulk Bi2212 identified an asymptotic 3D-XY critical region extending only
about $0.1\,$K above $T_c$, followed by a $0.15\,$K window of 3D Gaussian fluctuation, 
which are replaced by 2D Gaussian fluctuations at higher temperatures \cite{Monegotto97}. 

As a reference configuration we choose Bi2212 at 1~K above its $T_c$ with $z_c=\Delta r=5\xi$. 
This corresponds to a system with $\varepsilon\simeq 0.011$, measured by a pair of qubits placed 
about 40~nm above the surface and separated by 80~nm. Such scales are within reach of current 
experimental capabilities \cite{Le-subdiffraction,Degen-subtip,deLeon-subdiff}. 
However, for a film of one unit cell thickness, the equilibrium signal calculated from Eq. (\ref{eq:Seqlz}) 
for this geometry is only $4-8~\text{pT}^2/\text{Hz}$. Here, the lower and upper bounds represent independent 
versus perfectly correlated $\text{CuO}_2$ planes within the unit cell, respectively. 

To enhance the signal, one may utilize a stack of $N$ unit cells. 
Provided the cell thickness $c$ greatly exceeds the $c$-axis correlation length $c\gg\xi_c(T)$, 
the magnetic fluctuations in each cell contribute incoherently, and their intensities add linearly. 
Table \ref{tab:exp} indicates that for a Bi2212 film at $T=T_c$+1 K this condition is met  
since $\xi_c(T)=\xi_c(0)/\sqrt{\varepsilon}\simeq 0.1{\rm nm}\ll c=3.1{\rm nm}$. 
More importantly, as noted above, measurements of bulk Bi2212 show that the fluctuations become 
effectively two-dimensional, signaling the decoupling of different layers, already at $0.25$~K  
above $T_c$ \cite{Monegotto97}.
Under such conditions the exponential factor in the noise spectrum integral, 
Eq. (\ref{eq:S2deq}), becomes 
\begin{equation}
\label{eq:csum}
    \sum_{n=0}^{N-1}e^{-2\bar q (z_c+2nc)/\xi}=e^{-2\bar q z_c/\xi}\frac{1-e^{-4\bar q Nc/\xi}}{1-e^{-4\bar q c/\xi}}.
\end{equation}
For the case of interest $z_c\gg c$, and using the fact that the integral is dominated by $\bar{q} \lesssim \xi/z_c$, 
the denominator in Eq. (\ref{eq:csum}) may be approximated by $4\bar{q} c/\xi$. Consequently, one finds that 
\begin{equation}
    S_{zz}(z_c,N)=\frac{1}{2c}\int_{z_c}^\infty d\zeta\left[S_{zz}(\zeta,1)-S_{zz}(\zeta+2Nc,1)\right],
\end{equation}
where $S_{zz}(z_c,N)$ is the noise from a stack of $N$ layers. When $Nc\ll z_c$ this becomes 
$S_{zz}(z_c,N)=NS_{zz}(z_c,1)$. A more significant enhancement is achieved in the opposite limit 
$Nc\gg z_c$. Assuming that also $Nc\gg \Delta r$ and considering the far field $z_c\gg{\rm max}(\xi,c)$, 
the spatial dependence of the magnetic noise correlations changes from the forms derived above in 
Eqs. (\ref{eq:Seqlz}), (\ref{eq:E1signal}) and (\ref{eq:E2signal}) according to 
\begin{eqnarray}
\!\!\!\!\!\!\!\!\!&&O(E^0):\hspace{0.cm}  \frac{z_c}{(\Delta r^2+z_c^2)^{3/2}}\rightarrow 
\frac{1}{2c(\Delta r^2+z_c^2)^{1/2}}, \\
\!\!\!\!\!\!\!\!\!&&O(E^1):\hspace{0.cm}  \frac{\Delta r\, z_c}{(\Delta r^2+z_c^2)^{5/2}}
\rightarrow \frac{\Delta r}{6c(\Delta r^2+z_c^2)^{3/2}}, \\
\!\!\!\!\!\!\!\!\!&&O(E^2):\hspace{0.cm}   \left\{\renewcommand{\arraystretch}{2} 
\begin{array}{cc} 
\displaystyle{-\frac{1}{z_c^2}}\rightarrow -\frac{1}{2cz_c} & 
z_c\gg \Delta r  \\
\displaystyle{\frac{2}{\Delta r^2}}\cos2\theta\rightarrow  
\frac{1}{2c\Delta r}(\cos2\theta-1)& z_c\ll \Delta r
\end{array}\right. \!\!\!\!.
\renewcommand{\arraystretch}{1}   
\end{eqnarray}
For a 300~nm-thick Bi2212 film, a numerical evaluation, extending beyond the analytical 
approximations discussed above, yields an equilibrium signal of $90-180~\text{pT}^2/\text{Hz}$. 
As before, this range reflects the uncertainty regarding the degree of current correlation 
between $\text{CuO}_2$ planes within each unit cell. 

Turning to the non-equilibrium signal, we note that applying an identical, simultaneous pulse sequence 
$w(t)$ to both NV centers produces a filtering kernel in Eq. (\ref{eq:phasecov}) that is even in $\omega$. 
Consequently, synchronous identical phase-accumulation measurements are only sensitive to the 
frequency-even part of $S_{zz}(\omega)$, and therefore can not detect the $O(E)$ correction 
to the noise, Eq. (\ref{eq:SE1}). However, this limitation can be overcome by introducing a time delay  
to the second sensing sequence, such that $w_2(t)=w_1(t-\Delta t)$. This adjustment yields  
\begin{eqnarray}
\label{eq:phasecov2}
\nonumber
&&\!\!\!\!\!\!\langle\phi_1\phi_2\rangle=\gamma_{\rm NV}^2\int\frac{d\omega}{2\pi}|W(\omega)|^2 
\left[\cos(\omega\Delta t)S_R(\omega) \right.\\
&&\hspace{3.74cm}\left. -\sin(\omega\Delta t)S_I(\omega)\right],
\end{eqnarray}
where $S_R(\omega)=S_R(-\omega)$ and $S_I(\omega)=-S_I(-\omega)$ are the real, $\omega$-even and 
the imaginary, $\omega$-odd parts of $S_{zz}(\br_1,\br_2,\omega)$, respectively.

Our analysis shows that $O(E^2)$ non-equilibrium effects are expected to suppress the noise 
and introduce a $\cos2\theta$ component to the signal. The magnitude of these effects is 
governed by the factor $f^2(E,\varepsilon)$. Notably, the same properties that enhance the 
equilibrium noise in high-temperature superconductors, namely high $T_c$ and short $\xi$, 
simultaneously diminish $f^2(E,\varepsilon)$, as seen from Eq. (\ref{eq:fdef}) and 
Table \ref{tab:exp}. To estimate the maximum achievable non-equilibrium signal for a given film, 
we must consider the system's thermal budget. In the steady state, the Joule heating power 
per unit area is $P = E^2d/\rho$, where $d$ is the film thickness and $\rho$ is its resistivity. 
The resulting temperature rise is approximately $\Delta T = P(d/3\kappa_c+1/G)$, 
where $\kappa_c$ is the thermal conductivity in the direction perpendicular to the film and 
$G$ the thermal boundary conductance to the substrate. Consequently, the maximum 
field $E$ that can be applied for a given allowable $\Delta T$ scales as 
$E = \sqrt{\rho\Delta T (d^2/3\kappa_c+d/G)^{-1}}$.

The $c$-axis thermal conductivity of Bi2212 at 90~K is only $\kappa_c=0.67$~W~m$^{-1}$~K$^{-1}$ 
\cite{Zettl}. Lacking direct measurements for the thermal conductance of Bi2212 interfaces, we use 
a proxy value of $G=4\times10^8$~W~m$^{-2}$~K$^{-1}$, which is half the value reported for epitaxial 
SrRuO$_3$/SrTiO$_3$ oxide system \cite{GSTO}. Using $\rho=1.5~\mu\Omega~{\rm m}$ 
\cite{BSCCO-Aichner2025}, $d=60$~nm and $\Delta T=0.2$~K we determine the maximal 
applicable field to be $E=125$~V~cm$^{-1}$, yielding $f^2=0.126$. Numerical evaluation 
then predicts a mean noise reduction of $6.5-13~\text{pT}^2/\text{Hz}$, accompanied by a 
superimposed $\cos 2\theta$ component with an amplitude of $2-4~\text{pT}^2/\text{Hz}$. 
Higher signals can be achieved by pulsed driving. As an illustrative extrapolation of 
the quadratic result consider applying a field of 200 V cm$^{-1}$ 
to a $d=300$~nm film, which yields $f^2=0.33$. This increases the instantaneous amplitudes of the 
isotropic and quadrupolar components to $30-60~\text{pT}^2/\text{Hz}$ and $5-10~\text{pT}^2/\text{Hz}$, 
respectively. 

%Lowering the field to 100 V cm$^{-1}$ results in $f^2=0.083$, thereby placing the system deeper 
%in the perturbative regime, at the expense of lowering the associated signal by a factor of four.

Turning our attention to the signal expected from vortices, we use the parameter set of 
Ref. \cite{Curtis24} for a Bi2212 film to obtain $D=0.165$ cm$^2$/s. This appears to be a 
reasonable reference value, as the Bardeen-Stephen bare estimate is of order 1 cm$^2$/s \cite{Timm}. 
We note, however, that vortex motion in very thin Bi2212 films is thermally activated 
and strongly affected by pinning \cite{Rogers-diff}. Using this value gives $S_0^v/2\pi=9700$ 
pT$^2$/Hz. 

There is significant uncertainty regarding the value of the correlation length $\xi_+$, and consequently 
of the expected values for $\omega_v$ and $J_*$, in Bi2212 films. Numerical integration of the BKT RG equations in Ref. \cite{Curtis24} yields $\xi_+$ values ranging from 3~mm to 18~$\mu$m, as the temperature 
is raised from 1~K to 4~K above $T_{\rm BKT}$. Such large values may exceed the Pearl length, which is 
estimated to be around 60~$\mu$m in monolayer cuprates. In such a case our treatment should be modified 
to account for the effects of magnetic screening. Adopting a representative value of $\xi_+=10$~$\mu$m 
yields $\omega_v/2\pi\sim200$~KHz and $J_*\sim0.2$~A/m. 
Conversely, Corson {\it et. al.} used time-domain spectroscopy to 
measure the phase correlation time $\tau_\varphi$ in underdoped Bi2212 film \cite{Corson1999}. 
Converting their findings via $\xi_+\sim\sqrt{D\tau_\varphi}$ and the aforementioned value for $D$, results 
in $\xi_+\sim100$~nm near the onset of the isolated-layer BKT regime. This smaller scale implies  
$\omega_v/2\pi\sim$~GHz and $J_*\sim 17$~A/m. We note that the electric fields required for 
a sizable Gaussian $O(E^2)$ signal imply sheet currents per bilayer on the order of tens of A/m. 
Therefore, if vortex physics is at play current-induced vortex unbinding is expected to dominate under 
such driving conditions.

\section{Conclusion}

In this work, we have characterized the equilibrium and driven nonlocal magnetic noise generated by 
Gaussian superconducting fluctuations above $T_c$, with particular emphasis on the additional information 
provided by spatially separated magnetic probes. Within stochastic TDGL theory, the equilibrium signal is 
governed by the superconducting coherence length $\xi(T)$ and the Cooper-pair relaxation rate $\omega_0(T)$. 
In the far field, $z_c\gg\xi(T)$, the zero-frequency noise assumes a universal geometrical form, 
decaying as $z_c^{-2}$ for nearly coincident probes and as $\Delta r^{-3}$ at large lateral separation. 
Its frequency dependence remains essentially flat throughout the experimentally relevant NV sensing 
window before crossing over, on the much larger scale $\omega_0$, to an asymptotic $\omega^{-1}$ decay. Spatially resolved noise measurements therefore provide access not only to the overall strength of superconducting fluctuations but also to their characteristic length and time scales.

Driving the system out of equilibrium produces additional signatures that are absent from local 
equilibrium noise measurements. The correction linear in the electric field vanishes in the 
particle-hole-symmetric limit. Weak particle-hole asymmetry generates instead an $O(E)$ contribution 
that is odd in frequency and proportional to ${\bm E}\cdot\Delta{\bm r}$. Its explicit dependence on 
the probe separation makes this contribution intrinsically nonlocal. At $O(E^2)$, a finite quasi-static 
response exists even in the particle-hole-symmetric case. The drive suppresses the long-wavelength 
Gaussian magnetic noise and generates a quadrupolar $\cos2\theta$ anisotropy. These features reflect 
the anisotropic distortion of the fluctuating Cooper-pair spectrum by the applied field and provide 
direct information about the momentum structure of the nonequilibrium current correlations. A delayed 
two-sensor sensing protocol can, in addition, separate the frequency-odd $O(E)$ signal from the 
frequency-even contributions.

We have also developed a coarse-grained description of the magnetic noise produced by a free-vortex liquid 
above $T_{\rm BKT}$. Although Gaussian and vortex fluctuations exhibit a similar far-field spatial 
dependence in equilibrium, their characteristic scales differ substantially: the former are controlled 
by $\xi(T)$ and $\omega_0$, whereas the vortex response crosses over on the screening length 
$\kappa^{-1}\sim\xi_+$ and the much slower vorticity-relaxation scale $\omega_v=D\kappa^2$. The 
nonequilibrium response provides further distinctions. Opposite drift of vortices and antivortices 
couples fluctuations of their total density to the vorticity. At long wavelengths this channel enhances 
the static noise, in contrast to the suppression found for Gaussian fluctuations. On scales shorter 
than the screening length, however, drift-induced feedback provides an additional relaxation channel 
for vorticity fluctuations and generates a $-\cos2\theta$ quadrupolar component, corresponding to 
enhanced correlations for sensors separated perpendicular to the current. Thus both the sign and 
angular structure of the nonlinear response depend on whether the measurement resolves length scales 
shorter or longer than the vortex screening length.

At stronger drive, the distinction becomes richer because the applied current can itself unbind 
vortex-antivortex pairs. Above $T_{\rm BKT}$, this process becomes important once the current-induced unbinding scale falls below $\xi_+$. The resulting additional free vortices are expected to strengthen 
the short-distance, drift-induced $-\cos2\theta$ response, while simultaneously reducing the screening 
length. Consequently, the anisotropic vortex signal may evolve nonmonotonically with current as the 
experiment crosses from an effectively unscreened drift regime to a collectively screened one. 
This behavior is qualitatively different from the perturbative Gaussian response and provides another 
possible diagnostic of vortex-dominated fluctuations, although a quantitative treatment at $J\gtrsim J_*$ 
requires going beyond the fixed-density linearized vortex theory used here.

Our estimates indicate that cuprate films, and Bi2212 in particular, provide favorable conditions for 
observing these effects. Their short coherence length and high transition temperature enhance the 
equilibrium Gaussian signal, while multilayer films can further increase the measured noise through the 
incoherent addition of effectively decoupled CuO$_2$ layers. The nonlinear signal is more challenging 
because the electric fields required to make the $O(E^2)$ correction appreciable are constrained 
by Joule heating. The low $c$-axis thermal conductivity of Bi2212 makes internal heat flow through 
the film an important limitation in addition to the thermal boundary resistance. Thin films and 
pulsed or stroboscopic driving therefore appear particularly advantageous. At the same time, 
the currents required to obtain an appreciable Gaussian nonlinear signal can be comparable to or 
larger than the estimated vortex-unbinding scale, making the current dependence and angular 
structure of the noise especially valuable experimentally.

Taken together, these results show that nonlocal noise magnetometry can probe aspects of superconducting fluctuations that are largely inaccessible to a local noise measurement: their spatial correlations, directional response to a drive, characteristic relaxation scales, and nonlinear evolution with current. Consequently, combining measurements as a function of sensor height and separation, frequency, 
temperature, and current should therefore provide a means of discriminating Gaussian amplitude-and-phase 
fluctuations from vortex-dominated phase fluctuations in the regime above the superconducting transition.

\acknowledgements
We thank Nir Bar-Gill for useful discussions.

\renewcommand{\appendixname}{APPENDIX}
\appendix
\section{\MakeUppercase{Noise from a one-dimensional system}}

\begin{figure}[t!!!]
\centering
\includegraphics[width=\columnwidth,clip=true]{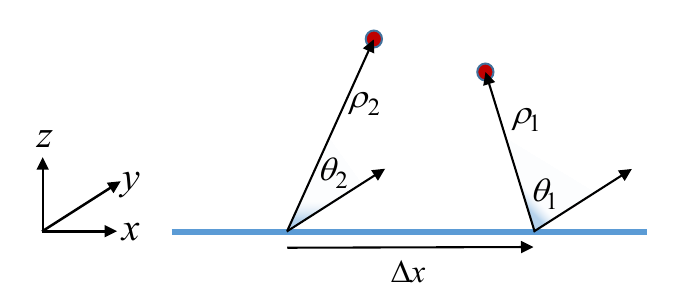}
\vspace{0pt}
\caption{The geometry of the one-dimensional system with the two NV centers.}
\label{fig:geometry1d}
\end{figure}

Consider a one-dimensional superconductor extending along the $x$ direction and sustaining an instantaneous 
electric current distribution $J(x,t)$. Neglecting retardation effects, the magnetic field is given by the 
Biot-Savart law (in Gaussian units)  
%we find for the magnetic field 
\begin{equation}
{\vec B}(\vr,t)=\rho\int dx' \frac{J(x',t)}{[(x-x')^2+\rho^2]^{3/2}}[0,-\sin\theta,\cos\theta],
\end{equation}
where the point $\vr=[x,\rho,\theta]$ is defined in the cylindrical coordinate system shown in Fig. \ref{fig:geometry1d}, 
while the field components are expressed in the Cartesian system. 
Specifically, the $z$ component is given by
\begin{equation}
\label{eq:Bz1d}
B_z(\vr,t)=2\int\frac{dq}{2\pi}\frac{d\nu}{2\pi}e^{i(qx-\nu t)}|q|K_1(|q|\rho)J(q,\nu)\cos\theta,
\end{equation}
where $K_1$ is the modified Bessel function. 

\subsection{Equilibrium noise}

A similar calculation to the one used in the two-dimensional case results in the current-current correlation function
\begin{equation}
\langle J(q,\nu)J(q',\nu')\rangle  = \frac{\pi^3 e^2T^2\xi}{T-T_c}I_{eq}(\bar q,\bar\nu)\delta(q+q')\delta(\nu+\nu'),
\end{equation}
where the one-dimensional equilibrium kernel is 
\begin{equation}
\label{eq:I1deq}
I_{eq}(\bar q,\bar \nu)= {\rm Re}\left[ \frac{1+\bar q^2+i\bar\nu-\sqrt{1+\bar q^2+2i\bar\nu}}
{\bar q^2\left(1+\bar q^2+2i\bar\nu\right)-\bar\nu^2}\right].
\end{equation}
%\begin{eqnarray}
%\label{eq:I1deq}
%\nonumber
%I_{eq}(\bar q,\bar \nu)&=&\Big[(\bar q +\bar q^3)^2-(1-\bar q^2)\bar\nu ^2\\
%\nonumber
%&&-{\rm Re}\left\{\left[ (\bar q^2-i\bar\nu)^2+\bar q^2\right]\sqrt{1+\bar q^2+2i\bar \nu}\right\}\Big]\\
%&&/\left[(\bar q^2 +\bar q^4+\bar\nu^2)^2-4\bar q^2\bar\nu^2\right]+O(\eta^2).
%\end{eqnarray}
Here, and throughout the equilibrium section, we set $\eta=0$. Consequently, we find
\begin{eqnarray}
\nonumber
\label{eq:S1deq}
S_{zz}^{eq}(\vr_1,\vr_2,\omega)&=&  S_0\frac{4}{\pi}\frac{T^2}{T_c^2}\int_{-\infty}^\infty d\bar q e^{i2\bar q\Delta x/\xi}I_{eq}(\bar q,\bar\omega)\\
\nonumber
&\times& \bar q^2K_1(2|\bar q|\rho_1/\xi)K_1(2|\bar q|\rho_2/\xi)\cos\theta_1\cos\theta_2,\\
\end{eqnarray}
where $\Delta x=x_1-x_2$.

\begin{figure}[t!!!]
\centering
\includegraphics[width=\columnwidth,clip=true]{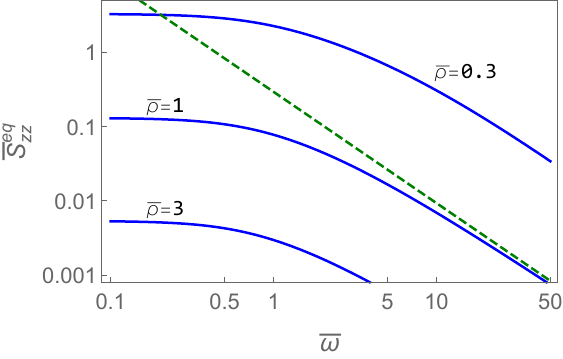}
\vspace{10pt}
\caption{$\bar S_{zz}^{eq}=S_{zz}^{eq}/[(T/T_c)^2S_0]$ as a function of $\bar\omega$ 
for the case $\vr_1=\vr_2$ and different values of $\bar \rho=\sqrt{\varepsilon}\rho/\xi_0$. 
The dashed line depicts a $\bar\omega^{-3/2}$ behavior.}
\label{fig:S1omega}
\end{figure}

\begin{figure}[t!!!]
\centering
\includegraphics[width=\columnwidth,clip=true]{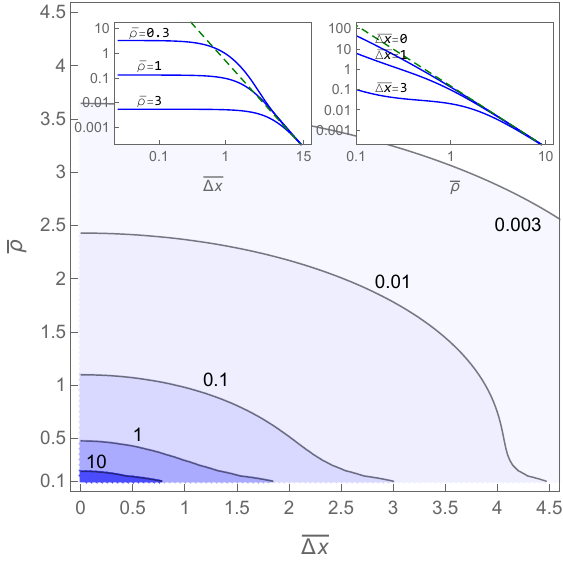}
\vspace{10pt}
\caption{Contour plot of $S_{zz}^{eq}(\omega=0)/[(T/T_c)^2S_0]$ for the case $\rho_1=\rho_2=\rho$ and $\theta_1=\theta_2=0$ 
as a function of $\bar \rho=\sqrt{\varepsilon}\rho/\xi_0$ and $\overline{\Delta x}=\sqrt{\varepsilon}\Delta x /\xi_0$. 
The left inset includes cuts as a function of $\overline{\Delta x}$, where the dashed line depicts a $\overline{\Delta x}^{\,-3}$ decay. 
The right inset shows cuts as a function of $\bar \rho$, where the dashed line represents a $\bar \rho^{\,-3}$ decay.}
\label{fig:S1xrho}
\end{figure}

Equations (\ref{eq:I1deq}) and (\ref{eq:S1deq}) provide a complete characterization of the magnetic noise. 
Figure \ref{fig:S1omega} illustrates its frequency dependence, which remains flat up to a characteristic 
crossover frequency of order $\omega_0$, followed by an asymptotic $\omega^{-3/2}$ decay. The dependence on 
sensor geometry is shown in Fig. \ref{fig:S1xrho}, where the signal exhibits $\rho^{-3}$ and $\Delta x^{-3}$ 
power-law decays in the far-field limit. In the following, we explore specific regimes where further 
analytical results can be derived.

\subsubsection{$\vr_1=\vr_2\equiv\vr$, $\omega=0$}
When only a single NV center is measured, the zero-frequency noise takes the form 
\begin{eqnarray}
\label{eq:eq1dsingleNV}
\nonumber
S_{zz}^{eq}(\vr,0)&=&-S_0\frac{2}{\pi}\frac{T^2}{T_c^2}\left[\frac{\pi^2\xi}{2\rho}+G_{24}^{41}\left(\frac{4\rho^2}{\xi^2}\left|
\begin{array}{cccc} 1/2 & \! & \! 1/2 & \!  \\ -1 & \!0 & \!0 &\! 1 \end{array}\!\! \right.\right)\right] \\
\nonumber
%&&-\left.\eta^2 G_{24}^{41}\left(\frac{4\rho^2}{\xi^2}
%\left|\begin{array}{cccc} -1/2 & \! & \! 1/2 & \! \\ -1 & \! 0 & \! 0 & \! %1\end{array}\!\!\right.\right)\right]
&&\times\cos^2\theta \\
&=&S_0\frac{T^2}{T_c^2}\cos^2\theta \left\{\renewcommand{\arraystretch}{2.5} 
\begin{array}{cc} 
\displaystyle{\frac{2}{\pi}
%(1+\frac{1}{\eta^2})
\frac{\xi_0^2}{\varepsilon\rho^2}} & \displaystyle{\rho\ll\frac{\xi_0}{\sqrt{\varepsilon}}} \\
\displaystyle{\frac{3\pi}{64}
%(1+\eta^2)
\frac{\xi_0^3}{\varepsilon^{3/2}\rho^3}} & 
\displaystyle{\rho\gg\frac{\xi_0}{\sqrt{\varepsilon}}}
\end{array}\right.
\renewcommand{\arraystretch}{1} 
,
\end{eqnarray}
where $G$ is the Meijer G-function. Note that for a clean one-dimensional system 
$\xi_0=1.3\,\xi_{0,\rm BCS}$, while in the dirty limit $\xi_0=1.5\,\sqrt{\xi_{0,\rm BCS}\ell}$ \cite{LarkinVarlamov}. 
%Here, and in the other equilibrium cases discussed below, we include the 
%leading dependence on the particle-hole symmetry breaking parameter $\eta$. 

\subsubsection{$\vr_1=\vr_2\equiv\vr$, $\rho\gg\xi_0/\sqrt{\varepsilon}$}
When the distance of the NV center from the wire significantly exceeds the coherence length, 
the exponential decay of the Bessel function $K_1(x \gg 1)$ ensures that the integral in 
Eq. (\ref{eq:S1deq}) is dominated by $|\bar{q}| \ll 1$. In this limit, we may expand
\begin{eqnarray}
\nonumber
I_{eq}(\bar q,\bar\omega)&=& I_{eq}(0,\bar\omega)+O(\bar q ^2)
= \frac{1}{\bar\omega ^2}{\rm Re}\left[\sqrt{1+2i\bar\omega}-1 \right]\\
%-\eta^2\left(\frac{1+i\bar\omega}{\sqrt{1+2i\bar\omega}}-1 \right)\right] 
&=&\left\{ \renewcommand{\arraystretch}{2.5} 
\begin{array}{cc} 
\displaystyle{\frac{1}{2}%(1+\eta^2)
-\frac{5}{8}%(1+3\eta^2)
\left(\frac{\omega}
{\omega_0}\right)^2} & \omega\ll \omega_0(T) \\
\displaystyle{%\left(1-\frac{\eta^2}{2}\right)
\left|\frac{\omega}{\omega_0}\right|^{-3/2}} &  \omega\gg \omega_0(T)
\end{array}
\renewcommand{\arraystretch}{1} 
\right. .
\end{eqnarray}
As a result,
\begin{equation}
S_{zz}^{eq}(\vr,\omega)=
S_0\frac{3\pi}{32}\frac{T^2}{T_c^2}\frac{\xi_0^3}{\varepsilon^{3/2}\rho^3}I_{eq}(0,\bar\omega)\cos^2\theta.
\end{equation}

\subsubsection{$\rho_1\gg\xi_0/\sqrt{\varepsilon}$, $\rho_1\gg\rho_2$, $\omega=0$}
Because $\rho_1\gg\xi$, we can still expand $I_{eq}(\bar q,0)\simeq 1/2$ in Eq. (\ref{eq:S1deq}). 
Furthermore, since $\rho_2\ll\rho_1$ we have that in the relevant range of momenta $|q|\rho_2\ll 1$ and we 
may use $K_1(|q|\rho_2)\simeq 1/(|q|\rho_2)$ to obtain
\begin{equation}
S_{zz}^{eq}(\vr_1,\vr_2,0)=S_0\frac{1}{4}\frac{T^2}{T_c^2}%(1+\eta^2)
\frac{\rho_1}{\rho_2}\frac{\xi_0^3}{[\varepsilon(\Delta x^2+\rho_1^2)]^{3/2}}\cos\theta_1\cos\theta_2.
\end{equation}

\subsubsection{$\rho_1=\rho_2\equiv\rho\gg\xi_0/\sqrt{\varepsilon}$, $\omega=0$}
Using the same considerations as above we find for this case
\begin{eqnarray}
\nonumber
&&S_{zz}^{eq}(\vr_1,\vr_2,0)\\
\nonumber
&&=S_0\frac{3\pi}{64}\frac{T^2}{T_c^2}\frac{\xi_0^3}{\varepsilon^{3/2}}%(1+\eta^2)
\cos\theta_1\cos\theta_2\frac{1}{\rho^3}\,_2F_1\left(\frac{5}{2},\frac{3}{2},2,-\frac{\Delta x^2}{4\rho^2}\right)\\
%&&S_{zz}(\vr_1,\vr_2,0)=S_0\frac{T^2}{T_c^2}\frac{\xi_0^3}{\varepsilon^{3/2}}%(1+\eta^2)
%\cos\theta_1\cos\theta_2\frac{\rho}{\Delta x^2(\Delta x^2+4\rho^2)^2}\\
%\nonumber
%&&\times\!\left[(\Delta x^2+4\rho^2){\cal K}\left(-\frac{\Delta x^2}{4\rho^2}\right) \right.
%+\left. (\Delta x^2-4\rho^2){\cal E}\left(-\frac{\Delta x^2}{4\rho^2}\right) \right]\\
\nonumber
&&= S_0\frac{T^2}{T_c^2}\frac{\xi_0^3}{\varepsilon^{3/2}}%(1+\eta^2)
\cos\theta_1\cos\theta_2\\
&&\hspace{0.43cm}\times\left\{  \renewcommand{\arraystretch}{3} 
\begin{array}{cc} 
\!\displaystyle{\frac{3\pi}{64}\frac{1}{\rho^3}\left(1-\frac{15\Delta x^2}{32\rho^2} \right)} & 
\rho\gg \Delta x \\
\!\displaystyle{\frac{1}{2|\Delta x|^3} + \frac{\rho^2\left[6\ln\left(\frac{2|\Delta x|}{\rho}\right)-11\right]}{2|\Delta x|^5} } &  
\rho\ll\Delta x
\end{array} 
\renewcommand{\arraystretch}{1} 
\right. .
\end{eqnarray}
%where ${\cal K}$ and ${\cal E}$ are the complete elliptic integrals of the first and second kind, respectively.

\begin{figure}[t!!!]
\centering
\includegraphics[width=\columnwidth,clip=true]{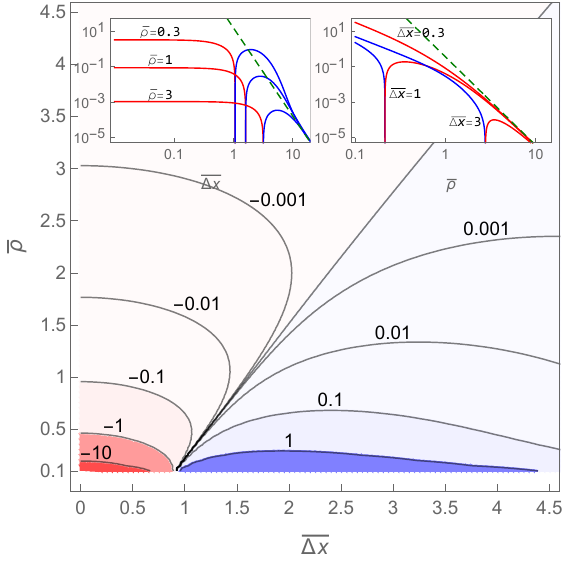}
\vspace{10pt}
\caption{Contour plot of $S_{zz}^{neq-ph}(\omega=0)/[S_0f_1^2(E,\varepsilon)(T/T_c)^2]$ for the case $\rho_1=\rho_2=\rho$ and $\theta_1=\theta_2=0$ 
as a function of $\bar \rho=\sqrt{\varepsilon}\rho/\xi_0$ and $\overline{\Delta x}=\sqrt{\varepsilon}\Delta x /\xi_0$. 
The left inset includes cuts of $|S_{zz}^{neq-ph}|$ as a function of $\overline{\Delta x}$, where the dashed line depicts 
a $\overline{\Delta x}^{\,-5}$ decay. The right inset shows cuts as a function of $\bar \rho$, where the dashed line 
represents a $\bar \rho^{\,-5}$ decay. Red (blue) segments correspond to negative (positive) values.}
\label{fig:S1neq}
\end{figure}

\subsection{Non-equilibrium noise}

In the presence of an electric field, the superconducting fluctuations contribute $\langle J(x,t)\rangle=\sigma_{AL} E$
to the average current density, with 
\begin{equation}
\sigma_{AL} = \frac{\pi e^2\xi_0}{16}(1+\eta^2)\frac{T}{T_c}{\frac{1}{\varepsilon^{3/2}}}.
\end{equation}
This average current produces a static magnetic field, which is of no concern to us here, but it 
also modifies the connected current correlations. To linear order in $E$, the modification vanishes 
identically when $\eta=0$. For $\eta\neq 0$ it becomes 
\begin{eqnarray}
\nonumber
\hspace{-0.9cm}&&\langle\langle J(q,\nu)J(q',\nu')\rangle\rangle_{neq}=\\
\hspace{-0.9cm}&&\frac{\pi^{5/2}}{2}f_1(E,\varepsilon)S_0\frac{T^2}{T_c^2}\frac{\xi_0^3}{\varepsilon^{3/2}}I_{neq}(\bar q,\bar\nu)\delta(q+q')\delta(\nu+\nu'),
\end{eqnarray}
where for the one-dimensional case we define
\begin{equation}
f_1(E,\varepsilon)=\frac{\sqrt{\pi}}{8}\frac{eE\xi_0}{T_c}\frac{1}{\varepsilon^{3/2}}.   
\end{equation}
Expanding $I_{neq}$ to lowest order in $\bar\nu$ and $\eta$ we find 
\begin{eqnarray}
\nonumber
&&I_{neq}(\bar q,\bar \nu)\equiv\eta\bar\nu I_{neq}(\bar q)=\\
\nonumber
&&\eta\bar\nu\frac{1}{\bar q^5(1+\bar q^2)^3}\left[\frac{4+23\bar q^2+4\bar q^4}{\sqrt{1+\bar q^2}}-4-21\bar q^2+6\bar q ^4-\bar q ^6\right].\\
\end{eqnarray}
In the case $\rho_1=\rho_2\equiv\rho\gg\xi$, which is dominated by $\bar q\ll1$, this leads to 
\begin{eqnarray}
\nonumber
&&S_{zz}^{neq}(\vec r_1,\vec r_2,\omega)=i\eta\frac{1575\pi^{3/2}}{2^{14}}S_0f_1(E,\varepsilon)\frac{T^2}{T_c^2}\frac{\xi_0^4}{\varepsilon^2}\frac{\Delta x}{\rho^5}\frac{\omega}{\omega_0}\\
\nonumber
&&\hspace{2.31cm}\times\,_2F_1\left(\frac{5}{2},\frac{7}{2},3,-\frac{\Delta x^2}{4\rho^2}\right)\cos\theta_1\cos\theta_2\\ 
\nonumber
&&=i\eta\frac{1575\pi^{3/2}}{2^{14}}S_0f_1(E,\varepsilon)\frac{T^2}{T_c^2}\frac{\xi_0^4}{\varepsilon^2}\frac{\omega}{\omega_0}\cos\theta_1\cos\theta_2\\
&&\hspace{2.31cm}\times\left\{  \renewcommand{\arraystretch}{2.5} 
\begin{array}{cc} 
\!\displaystyle{\frac{\Delta x}{\rho^5}} & 
\rho\gg \Delta x \\
\!\displaystyle{\frac{512}{15\pi}\frac{\rm sgn(\Delta x)}{\Delta x^4} } &  
\rho\ll\Delta x
\end{array} 
\renewcommand{\arraystretch}{1} 
\right. .
\end{eqnarray}

For the second order contribution to the connected correlations, we obtain in the case $\eta=0$ 
\begin{eqnarray}
\label{eq:neqJJ1d}
\nonumber
\langle\langle J(q,\nu)J(q',\nu')\rangle\rangle_{neq-ph}&=&\frac{\pi^5}{2^{11}} \frac{e^4 E^2 T^2\xi_0^3}{T_c^3}
\frac{1}{\varepsilon^{9/2}}I_{neq}^{ph}(\bar q,\bar\nu)\\
&&\times\delta(q+q')\delta(\nu+\nu'),
\end{eqnarray}
where the static non-equilibrium kernel is given by
\begin{eqnarray}
\label{eq:neqI1d}
\nonumber
&&\!\!\!I_{neq}^{ph}(\bar q,\nu=0)=\\
\nonumber
&&\;\;\;\;\!\!\!-\Big[16+68\bar q^2 +137\bar q^4+48 \bar q^6
+42\bar q^8+20 \bar q^{10}+5\bar q^{12} \\
&&\;\;\;\;\!\!\!-(16+60\bar q^2 +109 \bar q^4)\sqrt{1+\bar q^2}\Big]
/\left[\bar q^6(1+\bar q^2)^4\right].
\end{eqnarray}
%\begin{eqnarray}
%\label{eq:neqI1d}
%\nonumber
%I_{neq}^{ph}(\bar q,\nu=0)&=&\Big[16+76\bar q^2 +169\bar q^4+109 \bar q^6\\
%\nonumber
%&&-(16+68\bar q^2+137 \bar q^4+48\bar q^6 \\
%\nonumber&&+42\bar q ^8 +20 \bar q^{10}+ 5\bar q^{12})\sqrt{1+\bar q^2}\Big]\\
%&&/\left[\bar q^6(1+\bar q^2)^{9/2}\right].
%\end{eqnarray}
Equations (\ref{eq:neqJJ1d}) and (\ref{eq:neqI1d}) can be used together with Eqs. (\ref{eq:Szz}) and  (\ref{eq:Bz1d})
to numerically calculate the zero-frequency non-equilibrium noise. Representative results are shown in 
Fig. \ref{fig:S1neq}. We can  provide analytical expressions for the special cases considered above. 

\subsubsection{$\vr_1=\vr_2\equiv\vr$, $\omega=0$}
\vspace{-0.75cm}
\begin{eqnarray}
\nonumber
\hspace{-1cm}S_{zz}^{neq-ph}(\vr,0)&=&-S_0\frac{T^2}{T_c^2}f_1^2(E,\varepsilon)\cos^2\theta \\
&&\times \left\{ \renewcommand{\arraystretch}{2.5} 
\begin{array}{cc}
\displaystyle{\!\frac{103-30\pi}{15}\frac{\xi_0^2}{\varepsilon\rho^2}} & \;\; 
\displaystyle{\rho\ll\frac{\xi_0}{\sqrt{\varepsilon}}} \\
\displaystyle{\!\frac{4725\pi^2}{2^{17}}\frac{\xi_0^5}{\varepsilon^{5/2}\rho^5}} & \;\; 
\displaystyle{\rho\gg\frac{\xi_0}{\sqrt{\varepsilon}}}
\end{array}\right. 
\renewcommand{\arraystretch}{1} .
\end{eqnarray}
Note that the far-field scaling is different in the equilibrium case, Eq. (\ref{eq:eq1dsingleNV}), and 
the non-equilibrium setting considered here, while the near-field scalings match. 
Hence, unlike its two-dimensional counterpart, $f_1^2(E,\varepsilon)$ 
can be considered as a typical amplitude ratio between the near-field behavior in the two cases, rather than between their far-fields.

\subsubsection{$\vr_1=\vr_2\equiv\vr$, $\rho\gg\xi_0/\sqrt{\varepsilon}$}
In addition to the static result Eq. (\ref{eq:neqI1d}), we can extract the frequency dependence 
of $I_{neq}^{ph}(\bar{q}, \bar{\nu})$ for $|\bar{q}| \ll 1$. In this limit, which dominates the signal for 
large sensor heights, the noise spectrum becomes
%\begin{eqnarray}
%\nonumber
%S_{zz}^{neq}(\br,\omega)&=& \frac{\pi^4e^4E^2T^2}{T_c^3}\cos^2\theta\\
%&&\times\left\{ \renewcommand{\arraystretch}{2.8} 
%\begin{array}{cc} 
%\displaystyle{-\frac{4725}{2^{26}}\frac{1}{\varepsilon^{11/2}}\frac{\xi_0^5}{\rho^5}   
%+\frac{2097}{2^{20}}\frac{1}{\varepsilon^{9/2}}\frac{\xi_0^3}{\rho^3}\left(\frac{\pi}{32}\frac{\omega}{\varepsilon T_c}\right)^2} & \omega\ll T-T_c \\
%\displaystyle{\frac{15}{2^{19}}\frac{1}{\varepsilon^{9/2}}\frac{\xi_0^3}{\rho^3}\left(\frac{\pi}{32}\frac{\omega}
%{\varepsilon T_c}\right)^{-2}} &  \omega\gg T-T_c
%\end{array}
%\renewcommand{\arraystretch}{1} 
%\right. .\\
%\end{eqnarray}
%%\vspace{-0.8cm}
\begin{eqnarray}
\nonumber
\!\!\!\!\!\!\!\!\!\!&&S_{zz}^{neq-ph}(\vr,\omega)= \pi^2 S_0\frac{T^2}{T_c^2}f_1^2(E,\varepsilon)\frac{\xi_0^3}{\varepsilon^{3/2}\rho^3}\cos^2\theta\\
\!\!\!\!\!\!\!\!\!\!&&\times\left\{ \renewcommand{\arraystretch}{2.5} 
\begin{array}{cc} 
\displaystyle{-\frac{4725}{2^{17}}\frac{\xi_0^2}{\varepsilon\rho^2}   
+\frac{2097}{2048}\left(\frac{\omega}{\omega_0}\right)^2} & \omega\ll \omega_0(T) \\
\displaystyle{\frac{15}{1024}\left(\frac{\omega}
{\omega_0}\right)^{-2}} &  \omega\gg \omega_0(T)
\end{array}
\renewcommand{\arraystretch}{1} 
\right. \!.
\end{eqnarray}
\vspace{0.3cm}

%\vspace{1.8cm}
\subsubsection{$\rho_1\gg\xi_0/\sqrt{\varepsilon}$, $\rho_1\gg\rho_2$, $\omega=0$}
%%\vspace{-0.5cm}
\begin{eqnarray}
\nonumber
S_{zz}^{neq-ph}(\vr_1,\vr_2,0)&=&-\frac{315\pi}{512}\frac{T^2}{T_c^2}S_0f_1^2(E,\varepsilon)\frac{\xi_0^5}{\varepsilon^{5/2}}\\
\nonumber
\hspace{-1cm}&&\times\frac{\rho_1}{\rho_2}\frac{\rho_1^2-4\Delta x^2}{(\rho_1^2+\Delta x^2)^{7/2}}\cos\theta_1\cos\theta_2.\\
\end{eqnarray}

%\vspace{0.5cm}
\subsubsection{$\rho_1=\rho_2\equiv\rho\gg\xi_0/\sqrt{\varepsilon}$, $\omega=0$}
%%\vspace{-0.5cm}
\begin{eqnarray}
\nonumber
\!\!\!\!\!S_{zz}^{neq-ph}(\vr_1,\vr_2,0)&=&
\frac{315\pi}{512}\frac{T^2}{T_c^2}S_0f_1^2(E,\varepsilon)\frac{\xi_0^5}{\varepsilon^{5/2}}\cos\theta_1\cos\theta_2\\
&&\hspace{-2.5cm}\times\left\{  \renewcommand{\arraystretch}{3} 
\begin{array}{cc} 
\!\displaystyle{-\frac{15\pi}{256}\frac{1}{\rho^5}\left(1-\frac{35\Delta x^2}{16\rho^2} \right)} & 
\rho\gg\Delta x \\
\!\displaystyle{\frac{8}{|\Delta x|^5} + \frac{24\rho^2\left[5\ln\left(\frac{2|\Delta x|}{\rho}\right)-11\right]}{|\Delta x|^7} } &  
\rho\ll\Delta x
\end{array} 
\renewcommand{\arraystretch}{1} 
\right. .
\end{eqnarray}

\bibliography{magnoise}

\end{document}